# Phase-probability shaping for speckle-free holographic lithography

*Dong Zhao[#], Weiwei Fu[#], Ziqin Li, Jun He, Kun Huang[*]*

Department of Optics and Optical Engineering, University of Science and Technology of China, Hefei, Anhui 230026, China

[#] *D. Z. and W. F.* contributed equally to this work.

[*]Corresponding authors: K. H. (huangk17@ustc.edu.cn)

**Optical holography has undergone rapid development since its invention in 1948, but the accompanying speckles with randomly distributed intensity are still untamed now due to the fundamental difficulty of eliminating intrinsic fluctuations from irregular complex-field superposition. Despite spatial, temporal and spectral averages for speckle reduction, it is extremely challenging to reconstruct high-homogeneity, edge-sharp and shape-unlimited images via holography. Here we predict that holographic speckles can be removed by narrowing the probability density distribution of encoded phase to homogenize optical superposition. Guided by this physical insight, a machine-learning-assisted probability-shaping (MAPS) method is developed to prohibit the fluctuations of intensity in a computer-generated hologram (CGH), which empowers the experimental reconstruction of irregular images with ultralow speckle contrast ($C$=0.08) and record-high edge sharpness (~1000 mm$^{-1}$). It breaks the ultimate barrier of demonstrating high-end CGH lithography, thus enabling us to successfully pattern arbitrary-shape and edge-sharp structures such as vortex gratings and two-dimensional random barcodes.**

**Introduction**

Holography can reconstruct a predefined image from optical scattering of a randomly distributed phase or amplitude mask under the illumination of a high-coherence light source[1, 2]. The optical field at each position of the image plane can be taken as a coherent superposition of diffracted fields from all the pixelated sources at the holographic mask[3, 4]. The random phase or amplitude encoded in the mask introduces the uncertainty between constructive and destructive interferences[5-7], leaving bright and dark spots. Such fluctuations of intensity are named holographic speckles, which increase the noise level and destroy the uniformity, thereby constraining its applications in optical display[8-10] and lithography[11, 12]. Therefore, speckle reduction has become one of the central problems in optical holography towards practical usage.

Since randomness in the phase or amplitude mask is mandatorily required to reconstruct an irregular image, almost all the reported approaches for speckle reduction share the common origin of coherence control: the average of multiple incoherent speckles. For example, employing a degenerate-cavity laser with multiple orthogonal modes[13] and decreasing optical spatial coherence with a rotating diffusor[14] is taken as the spatial average of the speckles[15, 16], while the spectral average[17-19] is realized by broadening (or narrowing) the bandwidth of a laser[6] ( or a light-emitting diode[20]). In addition, the temporal average is also proposed by summarizing multiple speckles in a time sequence[21-23]. These averaging techniques can reduce the speckles efficiently but at the cost of blurred edges, missing details or time-consuming exposure. Subsequently, persistent efforts have been made by developing novel algorithms[20, 24, 25], but speckle reduction is achieved by introducing some additional technologies such as partially coherent sources[20] and camera-in-the-loop training[24], or by decreasing the edge sharpness of the target image with spatial filtering[25]. Therefore, it is still a fundamental challenge to remove the speckles in holographic reconstruction for a high-coherence laser, due to the difficulty in homogenizing the complex-field superposition of randomly phased light.

Here, we demonstrate that the speckle contrast of a holographic image can be theoretically suppressed down to nearly zero by shaping the probability density of the encoded random phase even under high-coherence illumination. This finding guides us to develop a machine-learning algorithm for hologram design with the phase-probability shaping method. Experimentally, we realize the designed phase by using dielectric geometric metasurfaces and reconstruct highly uniform and edge-sharp images (such as a binary "bull", a two-dimensional "barcode" and an optical "fork grating") to demonstrate the prototype of CGH lithography.

**Speckle suppression through phase-probability shaping**

First, we derive the statistical properties of optical speckles generated by optical diffraction from a random phase mask (Fig. 1a). If each phase $\varphi_n$ at the $n^{th}$ pixel in the mask is valued by obeying a Gaussian-shape probability of $P(\varphi) = exp[-\varphi^2/(2\sigma^2)]/(\sqrt{2\pi}\sigma)$ (where $\sigma$ determines its probability distribution[5]), the diffraction field at the position $(x, y, z)$ is expressed as $E=u+iv=Ae^{i\theta}$, where $u$, $v$, $A$ and $\theta$ are its real part, imaginary part, amplitude and phase, respectively. Because $\varphi_n$ is random by obeying the Gaussian-shape probability, $u$ and $v$ are also Gaussian random variables. When the number of the pixels is large, we use the central limit theorem[26] to calculate the $u/v$-joint probability density function, from which the intensity probability density $P_I(x,y,z,I)$ (where $I=A^2$) at position $(x,y,z)$ can be derived after several variable transformations (see Section 1 in Supplementary Materials). $P_I(x,y,z,I)$ is the local intensity probability density due to its strong dependence on the spatial position. After integration with respect to $x$ and $y$, we obtain the intensity probability density in the interested region $\Sigma$ of the speckle at a $z$-cut plane

$$P_I(z, I) = \frac{1}{S}\iint_\Sigma P_I(x, y, z, I)dxdy$$

$$= \frac{1}{S}\iint_\Sigma \left\{ \frac{1}{4\pi\sigma_u\sigma_v\sqrt{1-\rho_{u,v}^2}} \int_{-\pi}^{\pi} \exp\left[-\frac{\left(\frac{\sqrt{I}\cos\theta-\bar{u}}{\sigma_u}\right)^2 + \left(\frac{\sqrt{I}\sin\theta-\bar{v}}{\sigma_v}\right)^2 - 2\rho_{u,v}\cdot\frac{(\sqrt{I}\cos\theta-\bar{u})(\sqrt{I}\sin\theta-\bar{v})}{\sigma_u\sigma_v}}{2(1-\rho_{u,v}^2)}\right] d\theta \right\} dxdy, \quad (1)$$

where $S$ is the area of the region $\Sigma$, and the average of the variable $X$ is defined as $\bar{X} = \int X \cdot P(\varphi)d\varphi$, $\sigma_X^2 = \overline{(X-\bar{X})^2}$ and $\rho_{u,v} = \overline{(u-\bar{u})(v-\bar{v})}/(\sigma_u\sigma_v)$. Once $z$ and $\Sigma$ are fixed, the parameters $\bar{u}$, $\bar{v}$, $\sigma_u$, $\sigma_v$ and $\rho_{u,v}$ determine the relative probability density of intensity in the speckle. Since these parameters are the statistical average under the Gaussian probability $P(\varphi)$, Eq. (1) bridges the link between the statistical properties (*i.e.*, $P_I(z, I)$) of the speckle and the phase probability $P(\varphi)$ of the mask.

Now, we investigate the role of the mask's phase probability in determining the properties of the speckle. By using the Rayleigh-Sommerfeld integral[27], the simulated diffraction patterns of the masks with different phase- probability shapes (denoted by $\sigma$ in Fig. 1b) indicate stronger fluctuations of intensity for larger $\sigma$ (Fig. 1c). When $\sigma$ approaches infinity and $P(\varphi)$ is considered a uniform probability distribution, Eq. (1) is simplified into $P_I(z, I) = \frac{1}{S}\iint_\Sigma e^{-I/(2\sigma_u^2)}/(2\sigma_u^2)\, dxdy$ (where $\sigma_u^2 = \sum_{n=1}^N a_n^2/2$), which behaves like the Rayleigh statistics for a fully developed speckle[5]. Such an exponentially decaying probability density also exists in the diffracted patterns for $\sigma=0.6\pi$ and $0.7\pi$ (Fig. 1d) and exhibits nearly perfect agreement with the theoretical prediction by Eq. (1). Both the speckle contrasts[5] ($C = \sqrt{\overline{I^2}-(\bar{I})^2}/\bar{I}$, where the average variable $\bar{X} = \int X \cdot P_I(z,I)dI$) and the relative signal-to-noise ratios (*SNR*, which is defined as $1/C$) are close to 1, implying poor uniformity. In contrast, for the cases of small $\sigma$ (*e.g.*, $\sigma=0.1\pi$), the diffraction patterns are homogeneous with reduced fluctuations of intensity (Fig. 1c), leading to the narrowly distributed probability density (Fig. 1d) and high *SNR* (Fig. 1e). These results suggest that optical speckles can be suppressed by reducing the width of the phase probability distribution in the mask, which holds the physical origin of speckle-free holograms. Namely, the uniform probability distribution in the phase mask should be excluded during the hologram design.

**Hologram design with the MAPS method**

To design speckle-free holograms in a framework of phase-probability shaping, we develop the MAPS algorithm, as sketched in in Fig.s 2a and 2b and introduced in detail in the Methods. Our MAPS algorithm is built with three key features: 1) an iteration method based on the Adam gradient descent[28] to build up the phase-probability configuration by using the independent phase feedback of each pixel (Fig. 2b); 2) a hybrid cost function including *RMSE* (the root-mean-square error between the simulated and ideal images), *SD* (standard deviation of the image) and optical efficiency $\eta$ to fully evaluate the image quality; 3) the dynamic weighted coefficients $w_1$, $w_2$ and $w_3$ with their maximums controlled by the constants $w_{RMSE}$, $w_{SD}$ and $w_{Eff}$ (Fig. 2c) to control the singularity feature of the image.

By carrying out the MAPS algorithm with 1500 iterations, we designed different holograms by scanning the constants $w_{SD}$ and $w_{Eff}$ from 0 to 1 ($w_{RMSE}=1$ is used for all the simulations due to its dominating role in converging the diffraction pattern to the ideal image). To highlight the speckle features in Fig. 2d, we divide all the simulated images into two types: the patterns with (orange part) or without (cyan part) phase singularities (the simulated efficiency, RMSE and SD are discussed in Section 2 of Supplementary Materials). Both parts are separated by a linear boundary, which is the fitting of the discrete positions at the simulated boundary. In the parameter space of ($w_{SD}$, $w_{Eff}$), the cyan part has a larger area than the orange part, revealing that our MAPS algorithm can generate more singularity-free solutions than the singularity ones. To show their differences, two simulated diffraction patterns, labelled A ($w_{SD}=1$, $w_{Eff}=0.1$) and B ($w_{SD}=0.1$, $w_{Eff}=1$), are exemplified in Fig. 2e, revealing the good uniformity for case A and the dark spots with phase singularities for case B. These results have theoretically confirmed the validity of our MAPS algorithm in tailoring holographic speckles via the parametric operations of $w_{SD}$ and $w_{Eff}$.

Due to the different $w_{SD}$ and $w_{Eff}$ in both cases (A and B), their phase possibilities undergo completely distinguished evolution (Fig. 2f). For case A, the weight $w_2$ (because $w_{SD}=1$) dominates the cost function when $i<175$, where its phase probability, *RMSE* and efficiency (Fig. 2g) are unchanged but its *SD* decreases quickly down to zero to enhance the uniformity of the image. When $i>175$ (where the weight $w_1$ dominates), the phase probability density extends linearly with three peaks and then becomes stable for $i>400$ (where *RMSE* decreases monotonously for approaching the target but both the *SD* and efficiency go through one peak). In comparison, for case B, the initial $w_3$-dominated optimization expands the phase probability to enhance the optical efficiency, but the accompanying large *SD* leads to undesirable phase singularities that always exist in the following $w_1$-dominated optimization (because the phase probability is nearly unchanged).

Next, we reveal the fundamental importance of the phase probability in achieving the speckle-free holograms. Supplementary Figs. 1d and 1e show the designed phase profiles for cases A and B, indicating a spatially dependent distribution of phase probability. Inside the "bull"-like contour (caused by the short imaging distance of Fresnel hologram), the phase probabilities for both cases have the similar Gaussian distributions (*i.e.*, their overlying peaks located between 0 to $\pi/2$ in Fig. 2h) to reconstruct the expected image. But, outside the "bull"-like region, case B has a uniformly distributed phase probability, which leads to a fully developed speckle according to the prediction by Eq. (1). The interference between this speckle and the reconstructed image yields the isolated phase singularities with dark centers (Fig. 2e, the lower panel). For case A, the phase probability outside the "bull"-like region has a narrower distribution with two Gaussian peaks (see Fig. 2h), resulting in much lower speckle that can be removed through the optimization (Fig. 2e, the upper panel). The above analysis indicates that the speckle-free hologram designed by machine learning can be well-understood from the viewpoint of phase-probability shaping. The issues about diffraction efficiency and operating conditions of the proposed speckle-free hologram are provided in Section 2 of Supplementary Materials.

**Experimental verification of speckle-free holograms**

To verify these predictions, two exemplified phase profiles are designed by using the MAPS (Fig. 3a) and GS (Fig. 3b) algorithms, respectively. For our MAPS case, the phase probability density has a narrow distribution (featured by the zero probability at -0.86$\pi$, Fig. 3c), which allows less fluctuations of intensity than that (taken as a superposition of the uniform and Gaussian shapes) for the GS case. As expected, the simulated image (Fig. 3d) for the MAPS case reveals better uniformity (with the speckle contrast $C = SD/\bar{I} = 0.07$) than that ($C$=0.46) of the GS case (Fig. 3e).

Experimentally, we utilize geometric metasurfaces[29-34] composed of rotating dielectric nanobricks (see the design details in Section 3 of Supplementary Materials) to realize phase masks, which are fabricated via standard electron-beam lithography with a wet-etching process (see the fabrication steps in Section 4 of Supplementary Materials). The well-fabricated samples (see their scanning-electron-microscopy (SEM) images in Figs. 3f and 3g) enable us to achieve high-fidelity holographic images in Figs. 3h and 3i by using a self-built measurement setup (Section 5 in Supplementary Materials). By comparing their experimental speckle contrasts ($C$=0.08 and $C$=0.50 for the MAPS and GS cases respectively) and line-scanning intensity profiles (Fig. 3j), we confirm the expected speckle reduction by using our MAPS algorithm. To ensure this further, the probability density of its normalized intensity is calculated with a FWHM (*i.e.*, full width at half maximum, Fig. 3k) of 0.122, which is approximately one order of magnitude smaller than that (FWHM=1.14) for the GS image. A further discussion about the experimental factors influencing the uniformity of the holographic images is provided in Section 6 of Supplementary Materials.

Due to the dispersion-less properties of geometric metasurfaces[35], the speckle-free features are valid over a wide spectrum (see the measured images in Supplementary Fig. 3). Nearly constant values near $\lambda$=633 nm are observed in the simulated and experimental FWHMs (Fig. 3l) and speckle contrasts (Fig. 3m). For $\lambda$<608 nm, the absorption of silicon in geometric metasurfaces decreases the conversion efficiency (Supplementary Fig. 4) from the incident circular polarization to its cross polarization, thus increasing the background that leads to fluctuations of intensity (featured by larger FWHMs and speckle contrasts as shown in Figs. 3l and 3m). Nevertheless, the MAPS case still has better uniformity over the broadband spectrum than the GS case.

To characterize the edge sharpness, we employ the inverse of the spatial range between the 10% and 90% of the maximum intensity at the edge[13] (Fig. 3n). Figure 3o shows the simulated and measured sharpness of ~1000 mm$^{-1}$, which is enhanced by nearly ~100 times compared with the latest reported sharpness via the spatial-mode average[13]. Such a giant enhancement originates fundamentally from our high-coherence and single-mode reconstruction via phase-probability shaping, where none of the previous averaging or filtering methods is included. Moreover, by using larger-size metasurfaces that offer much higher spatial frequencies[3, 36] for holographic reconstruction, we can enhance the edge sharpness further to fabricate finer patterns in optical lithography.

**Prototype of holographic lithography**

To demonstrate proof-of-concept holographic lithography, we design a phase mask at $\lambda$=405 nm and fabricated relative geometric metasurfaces made of less-absorption silicon-nitride nanobricks (see the SEM image in Fig. 4a). To show the capacity of patterning complex structures, two holographic images (*i.e.*, a two-dimensional random barcode and a fork grating) are exemplified with their measured intensity profiles shown in Figs. 4b and 4c, respectively. Due to the limited imaging distance of 390 μm, the holographic images are magnified first by using a couple of objective lenses and then projected onto the photoresist (Microposit S1813, Shipley) with an exposure time of 10 seconds (see the details in Section 7 of Supplementary Materials).

The developed photoresist patterns (Figs. 4d and 4e) show the expected details with clear edges, implying highly homogeneous exposure. In the exposure area, one cannot see random speckle dots or broken lines, which appear in previous attempts [11, 12, 37, 38] to demonstrate holographic lithography due to their un-eliminated speckles. Although the smallest features are tens of micrometers achieved in a laboratory environment, the resolving accuracy of lithography can be enhanced by using larger-sized metasurfaces with a long working distance that enables direct exposure without projection objectives. According to Fourier optics[3], its theoretical resolution limit is ~0.5$\lambda$/*NA* (*NA* is the numerical aperture of the hologram) for a single exposure.

**Discussion**

Such holographic lithography is doubly examined by characterizing the optical performance of the patterned fork grating. Under the illumination of laser light (see Section 8 in Supplementary Materials), the fork grating diffracts light into different orders, where the dark centers in ±1-orders (Fig. 4f) indicate the broadband creation of the expected vortex phase[39, 40]. To quantitatively evaluate its performance, the photoresist in the fork grating is measured with a height of *h*=950 nm (Fig. 4g) by using a profilometer, yielding a wavelength-dependent phase delay (Fig. 4h) of $\psi=2\pi(n-1)h/\lambda$ (*n* is the refractive index of the photoresist). When such phase delay is odd times of $\pi$, the ratio of the 1st-order intensity to the total intensity of 0 and ±1 orders reaches its maximum of 0.5, which is verified by the simulated and experimental results with high consistency (Fig. 4h). These results confirm that, this fork grating becomes the first optics-level and high-quality element delivered by using CGH lithography. Particularly, it is important to distinguish this CGH lithography with the interference-based holographic lithography[41] that can only pattern the periodic structures.

In summary, we have reported high-uniformity, edge-sharp and shape-unlimited holographic images under the illumination of single-mode, high-coherence lasers by shaping the phase-probability with the physics-guided MAPS algorithm. Benefiting from these high-quality images, we have presented a prototype of computer-generated-holography lithography that exhibits the capability of patterning optics-level complex microstructures in parallel. Although static metasurfaces are used in this holographic lithography, a spatial light modulator (SLM) carrying the phase designed via our MAPS algorithm is also able to dynamically reconstruct these speckle-free images (not shown), which can be scaled down for higher-resolution patterning. The reconfigurable feature of SLM allows multiple mask-less exposures without any alignment error, thus promising fast, less-expensive and high-accuracy fabrication. Therefore, this technique offers the possibilities to develop low-cost, parallel, dynamic, and mask-less lithography for fabricating large-size nanophotonic devices efficiently.

**Methods**

**Machine-learning-assisted probability-shaping algorithm**. This algorithm is developed for designing the phase-type compute-generated hologram with the controllable probability-shaping method. Here, the geometric metasurfaces are used to realize an arbitrarily designed phase (twice the rotating angle of the dielectric nanobricks[32]) for the incident circular-polarization light because their subwavelength pixel pitches can exclude the undesired twin images and high diffraction orders[42, 43]. Figure 2a sketches the reconstruction of the holographic image through optical diffraction from phase-encoded geometric metasurfaces. The design algorithm contains the following key steps:

I. Simulating the diffraction pattern. The rigorous Rayleigh-Sommerfeld diffraction integral[27] is employed to numerically simulate the diffraction patterns without any approximation. In each iteration, the simulated intensity pattern at the target plane is employed to calculate the key parameters such as *RMSE*, *SD* and optical efficiency.

II. Updating the key parameters. To avoid any confusion, we provide their definitions: $RMSE = \sqrt{\sum_{n=1}^{N}(I_n/\bar{I} - I_{ideal})^2/N}$, $SD = \sqrt{\sum_{m=1}^{M}(I_m - \bar{I})^2/M}$, $\eta = P_{image}/P_{incident}$, where $I_n$ is the intensity at the $n^{th}$ pixel located in the entire simulated pattern (containing *N* pixels), $\bar{I}$ is the average intensity only at the image region, $I_m$ is the intensity at the $m^{th}$ pixel located only in the image (containing *M* pixels), $I_{ideal}$ is the intensity of the ideal target, $P_{image}$ is the total power encircled within the image and $P_{incident}$ is the total power of incidence. By using the above definitions, we can update the *RMSE*, *SD* and optical efficiency in one iteration. Note that, these three parameters are highly dependent on the simulated patterns and hence become the key parameters to evaluate the image quality of the hologram. The RMSE can evaluate the convergence of the simulated image to the ideal one, the SD is critical in removing the phase singularities with dark intensity and enhancing the uniformity of the holographic image, and the parameter $\eta$ is employed to improve the optical efficiency of the hologram.

III. Constructing the cost function. To obtain holographic images with high quality such as good uniformity without any speckle and high efficiency, we adopt a multi-parameter cost function with the form of $CF_i = w_1(i) \cdot RMSE_i + w_2(i) \cdot SD_i + w_3(i) \cdot (1 - \eta_i)$, where $i$ is the index of the iterations, and $w_1(i)$, $w_2(i)$ and $w_3(i)$ are the $i$-dependent weights of *RMSE, SD* and $\eta$, respectively. The reason we choose these three parameters with $i$-dependent weights is threefold. First, the commonly used *RMSE* in various holographic algorithms is responsible for only the similarity between the simulated and ideal images, so the uniformity and efficiency issues concerned here are not well-solved by using a single *RMSE* parameter. This issue has been observed in many reported algorithms. Second, both the SD and efficiency also have more serious singleness of functionality so that they are seldom used in most algorithms. If the weight of SD is too large in a fixed cost function, the designed image might have low efficiency despite good uniformity; a similar issue occurs for the parameter $\eta$ of efficiency. Third, a dynamic cost function is employed by using the dynamic weight witha strong dependence on the iteration. To realize the dynamic cost function, our strategy of arranging these three parameters is that, the *SD* and efficiency parts dominate only the beginning of the iterations while the *RMSE* is the only parameter in the cost function for the remaining iterations. Following this strategy, we suggest $w_1(i) = w_{RMSE}\{tanh[(i - i_0)/D_0] + 1\}/2$, $w_2(i) = w_{SD}\{tanh[-(i - i_0)/D_0] + 1\}/2$, and $w_3(i) = w_{Eff}\{tanh[-(i - i_0)/D_0] + 1\}/2$, where $w_{RMSE}$, $w_{SD}$ and $w_{Eff}$ are used to control the weight of each parameter in the cost function, $i_0$ is the transition position between two states in the iterations, and $D_0$ determines the length of the transition process. Here, $i_0$=175 and $D_0$=40 are used in this work to offer a nearly null *SD* for $i<i_0$ (see case A in Fig. 2g) so that we can exclude any solutions that lead to the phase singularities in the holographic images. Their dependences on the iterations from 1 to 500 are shown in Fig. 2c for a better understanding of its working principle. The constants $w_{RMSE}$, $w_{SD}$ and $w_{Eff}$ are important in investigating their roles in phase-probability shaping and enhancing the image quality. Note that, $w_{RMSE}$=1 is employed throughout our optimization because of the *RMSE*'s good performance in optimizing the hologram. However, $w_{SD}$ and $w_{Eff}$ are valued between 0 and 1 to investigate the tradeoff between the uniformity and efficiency for the hologram design.

IV. Independent phase feedback via the gradient descent method of Adam. The Adam method[28] is used here to offer an independent increment of each phase so that we can shape the phase probability density by tuning the dynamic cost function. At position ($x_n$, $y_n$) in the $i^{th}$ iteration, the gradient of the cost function with respect to the local phase $\varphi(x_n, y_n)$ is defined as $g_i = \partial(CF_{i-1})/\partial\varphi_{i-1}(x_n, y_n)$, which is realized numerically by calculating the changing value of *CF* under a small variation of $\varphi(x_n, y_n)$ (keeping all other phase values unchanged). To update all the phase values in one iteration, *N* (*i.e.*, the total number of all the pixels) gradient calculations are needed so that the cost in both time and computing is extremely high for a large *N*. Therefore, our algorithms are implemented in an open-access TensorFlow for accelerating the optimization. Based on the calculated gradient $g_i$, we can update the phase (see Fig. 2b) $\varphi_i(x_n, y_n) = \varphi_{i-1}(x_n, y_n) - \alpha \hat{m}_i/(\sqrt{\hat{v}_i} + \varepsilon)$ , where $\hat{m}_i = [\beta_1 m_{i-1} + (1-\beta_1)g_i]/(1-\beta_1^i)$, $\hat{v}_i = [\beta_2 v_{i-1} + (1-\beta_2)g_i^2]/(1-\beta_2^i)$, $\beta_1$=0.9, $\beta_2$=0.999, $\varepsilon$=10$^{-8}$, $m_0$= $v_0$=0 and $\alpha$=0.001. Note that, each phase $\varphi_i(x_n, y_n)$ is updated independently, which enables this algorithm to shape the phase probability. Such phase-probability shaping can be observed in different examples, as shown in Figs. 2f and 2g. This feature in our algorithm is not available in other algorithms that are based on the inverse propagation of the target image, such as the Gerchberg-Saxton algorithm[44] and Yang-Gu algorithm[45].

The above four steps are the core of our hologram design method. In fact, the Adam approach in our algorithm offers the entire configuration of phase-probability shaping, while the weights $w_{RMSE}$, $w_{SD}$ and $w_{Eff}$ are the key to controlling the image quality of the hologram. To test its performance, we run the code on a personal computer (CPU Intel Core i5-7500, 32G RAM). The details of the parameters are the pixel pitch px=py=250 nm, the hologram size of 128 μm×128 μm, the propagation distance of z=250 μm between the metasurfaces and the image plane, and the operating wavelength λ=633 nm. Under such conditions, it takes ~42 minutes to run 1500 iterations.

**Acknowledgement**

K.H. thanks the National Key Research and Development Program of China (No. 2022YFB3607300), the National Natural Science Foundation of China (Grant Nos. 12134013 and 61875181), CAS Pioneer Hundred Talents Program, CAS Project for Young Scientists in Basic Research (Grant No.YSBR-049) and the support from the University of Science and Technology of China's Centre for Micro and Nanoscale Research and Fabrication.

**Author contributions**

K. H. conceived the idea and developed the theory. W. F., J. H. and K. H. performed the simulations. D. Z. and Z. L. prepared and fabricated optical samples. W. F. and D. Z. built up the experimental setup and performed the characterization. K. H. and D. Z. wrote the manuscript. K. H. supervised the overall project. All authors discussed the results, carried out the data analysis and commented on the manuscript.

**Competing financial interests**

The authors declare no competing financial interests.

**Figures and captions**

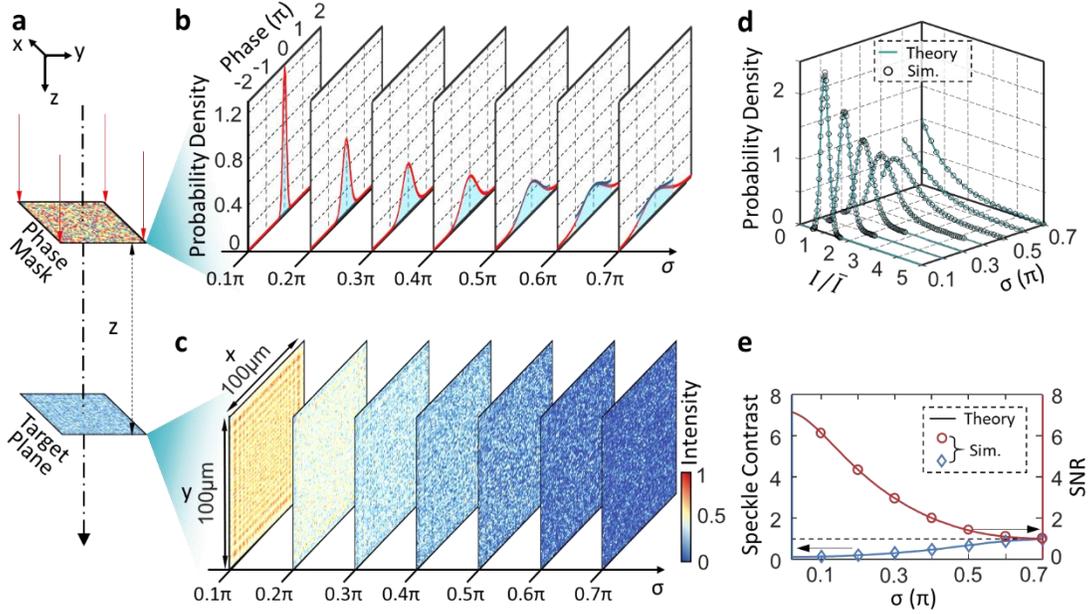

**Figure 1. Working principle of speckle suppression through phase-probability shaping**. **(a)** Sketch for optical diffraction from a random phase mask that is located at *z*=0. The size of the phase mask is exemplified to be 128 μm×128 μm, resulting in the pixel number *N*=512×512 because $\Delta x=\Delta y=0.25$μm in our simulations. The propagation distance between the phase mask and the target plane (where the speckle is evaluated in this work) is chosen at *z*=250 μm for the proof-of-concept demonstration. **(b)** Probability density distributions of the phase mask. The red curves show the probability shapes that are labelled by different standard deviations from $\sigma=0.1\pi$ to $\sigma=0.7\pi$ with an interval of $0.1\pi$. The cyan parts exhibit equivalent shapes after phase wrapping for $|\varphi|>\pi$. **(c)** Simulated diffraction patterns from the phase masks with different $\sigma$. The region of interest $\Sigma$ is a square of 100 μm×100 μm centered on the position (*x*=0, *y*=0, *z*). For a better comparison, all intensity profiles are normalized to the maximum intensity of the $\sigma=0.1\pi$ case. **(d)** Theoretical (solid curves) and simulated (black hollow circles) probability density distributions of the diffraction patterns at *z*=250 μm. The theoretical results are obtained by using Eq. (1). **(e)** Speckle contrasts (left) and the signal-to-noise ratios (right). The solid curves denote the theoretical prediction by Eq. (1), while the diamonds and circles are obtained by the statistical analysis of the diffraction patterns in **(c)**.

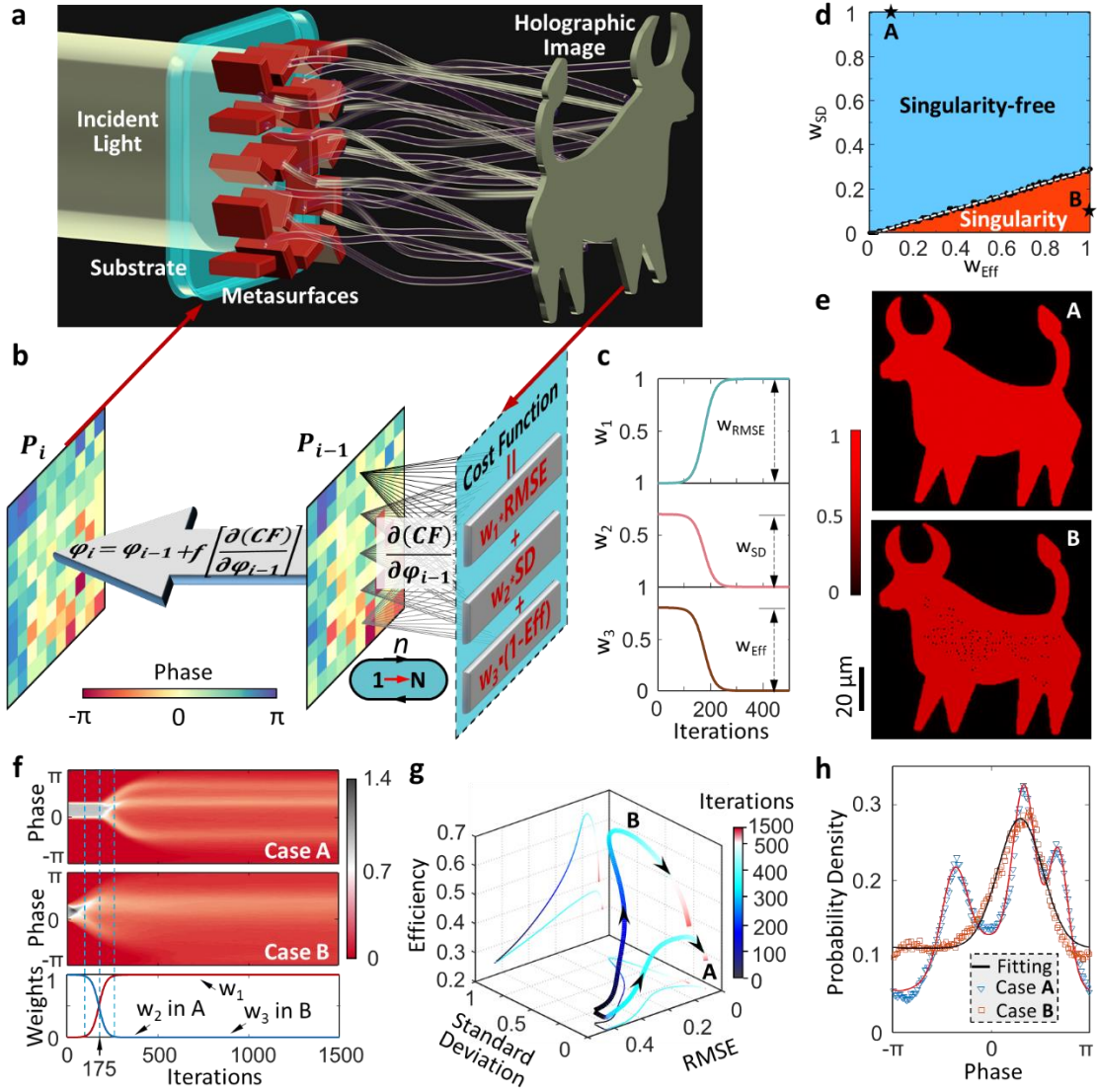

**Figure 2. Hologram design via phase-probability shaping.** (**a**) Sketch for holographic imaging using geometric metasurfaces. (**b-c**) Flowchart (**b**) for the phase updating strategy in each iteration. The Adam gradient descent is used in the hologram optimization. The cost function is a weighted superposition of RMSE, standard deviation and optical efficiency, where their weights $w_1$, $w_2$ and $w_3$ depend tightly on the index (labelled by $i$) of the iterations, as shown in (**c**). The mathematical expressions of these weights are provided in the Methods. (**d**) Map of phase singularity in the optimized images for different $w_{SD}$ and $w_{Eff}$. Depending on both parameters (because $w_{RMSE}=1$), the phase singularity disappears (cyan) or exists (orange) in the images. Both parts have the boundary of a white dashed line ($w_{SD}=0.2956w_{Eff}-0.008096$), which is obtained by fitting our simulated boundary (black dots). In our simulations, a metasurface hologram with a pixel pitch of 250 nm × 250 nm is illuminated by a laser light of $\lambda=633$ nm and then reconstructs a "bull"-pattern image at a propagating distance of $z=250$ μm, which is chosen to match

the simulations in Fig. 1. **(e)** Simulated images without (A) or with (B) phase singularities. **(f)** Probability evolution of the designed phases for the singularity-free (top) and singularity (middle) cases. Their weights with increasing iterations (bottom) are also plotted for a better observation of the phase-probability change. **(g)** Parametric (*i.e.*, optical efficiency, SD and RMSE) evolution of the optimized image with the increment of the iteration number (distinguished by the line color). Cases A and B are the evolution processes for a holographic image with or without phase singularity. Their parametric evolutions go along two completely different paths, finally arriving at (*RMSE*=0.025, *SD*=0.079, $\eta$=0.296) for A and (*RMSE*=0.038, *SD*=0.197, $\eta$=0.396) for B. Despite good uniformity, case A has lower efficiency than case B, which reveals a fundamental trade-off between uniformity and efficiency for all holographic designs. **(h)** Phase probability distributions of designed masks for the singularity-free (triangles) and singularity (squares) cases after 1500 iterations. For case A, the fitted probability density (red curve) has the form of $P(\varphi) = \sum_{m=1}^{4} a_m exp\left[-\left(\frac{\varphi-b_n}{c_m}\right)^2\right]$, where $a_1$=0.2141, $a_2$=0.1658, $a_3$=0.1181, $a_4$=0.1249, $b_1$=1.067, $b_2$=2.15, $b_3$=-1.169, $b_4$=0.15, $c_1$=0.5, $c_2$=0.45, $c_3$=0.57 and $c_4$=2.55. For case B, the fitted probability density (black curve) is $P(\varphi) = 0.1106 + 0.171 \times exp\left[-\left(\frac{\varphi-0.935}{1.007}\right)^2\right]$. In the fitted results, the phase probability for $|\varphi|>\pi$ is wrapped into the region $|\varphi|\leq\pi$.

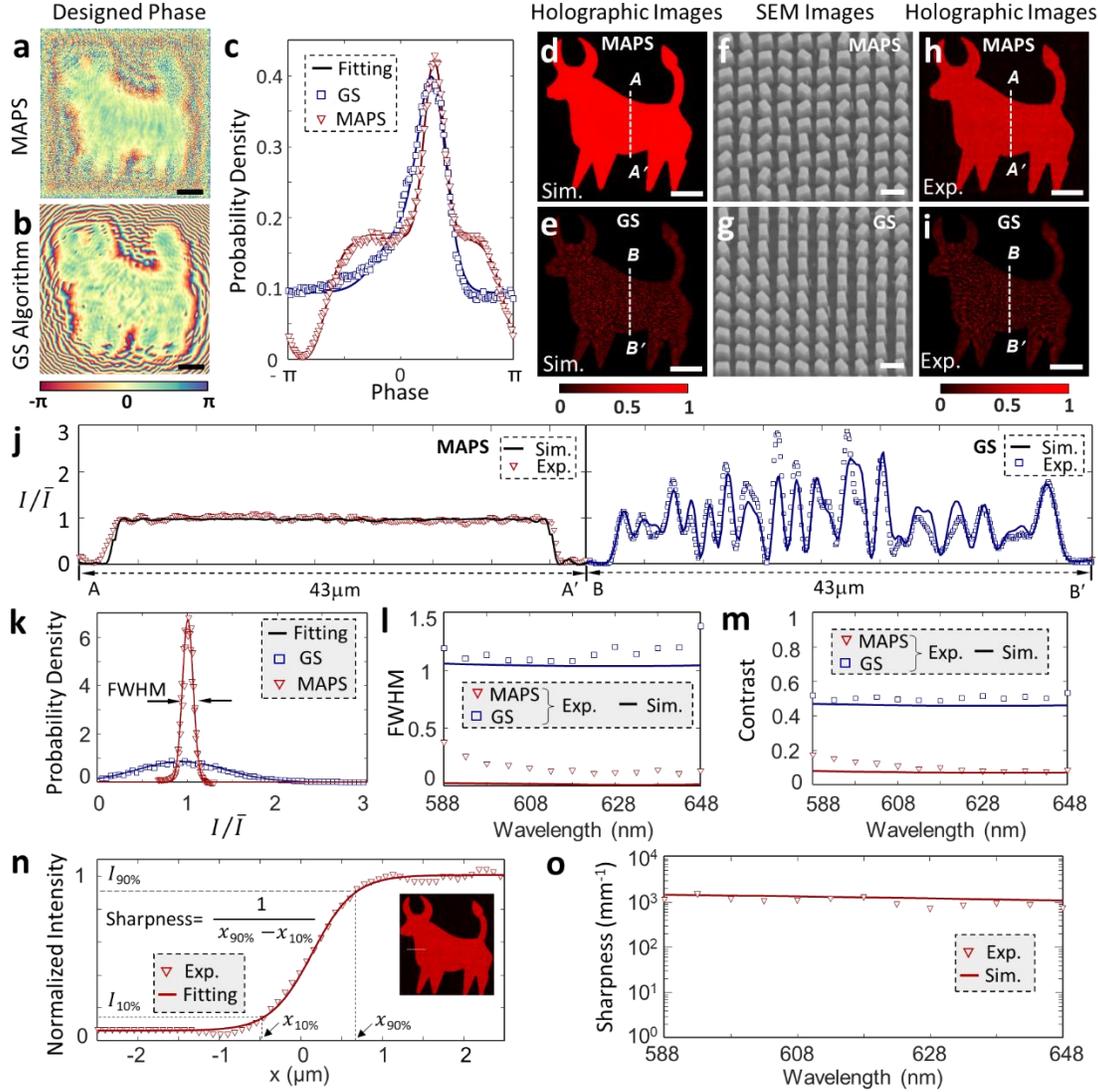

**Figure 3. Experimental characterization of speckle-free holograms.** (a-b) Designed phase profiles by using MAPS (a) and GS (b) algorithms. Without the loss of generality, we use the parameters $w_{SD}=0$ and $w_{Eff}=0$ (near the boundary but still in the singularity-free region of Fig. 3(d)), so that the designed holographic images have no speckle. All the other parameters in both the MAPS algorithm and optical system are identical to those in Fig. 2. (c) Probability density distributions of the designed phases in (a) and (b). The MAPS case has a fitted probability density of $P(\varphi) = 0.2504 \times exp\left[-\left(\frac{\varphi-4.067}{0.3872}\right)^2\right] +$ $0.1714 \times exp\left[-\left(\frac{\varphi-3.55}{2.557}\right)^8\right]$ (red curve), while $P(\varphi) = 0.0949 + 0.0875 \times exp\left[-\left(\frac{\varphi-3.233}{1.003}\right)^2\right] +$ $0.2557 \times exp\left[-\left(\frac{\varphi-4.028}{0.5478}\right)^2\right]$ for the GS case (blue curve). (d-e) Simulated intensity profiles for the MAPS (d) and GS (e) cases. The dashed lines (*AA′* and *BB′*) denote the position for line-scanning intensity for a comparison between the simulated and experimental results. Scale bars: 20 μm. (f-g)

SEM images of the fabricated silicon metasurfaces for the MAPS (**f**) and GS (**g**) cases. Scale bars: 300 nm. (**h-i**) Measured intensity profiles for the MAPS (**h**) and GS (**i**) cases. Scale bars: 20 μm. (**j**) Comparison of the experimental and simulated line-scanning intensity profiles for the MAPS (left panel, along *AA'* in (**d**) and (**h**)) and GS (right panel, along *BB'* in (**e**) and (**i**)) cases. (**k**) Probability density distributions of measured intensity profiles for the MAPS (triangles) and GS (squares) cases. The Gaussian shape is used to fit the experimental probability density. (**l-m**) The FWHMs (**l**) of the intensity probability density and speckle contrasts (**m**) at the different wavelengths. The simulated data are shown as solid lines and the experimental data are labelled by triangles for the MAPS and squares for the GS case. (**n**) Measured (curve) and fitted (line) line-scanning intensity profile at the edge of the holographic image. The insert shows the edge position of the experimental image. The sharpness is calculated by using the formula in the insert. (**o**) Experimental (triangles) and simulated (curve) sharpness of the images at the different wavelengths.

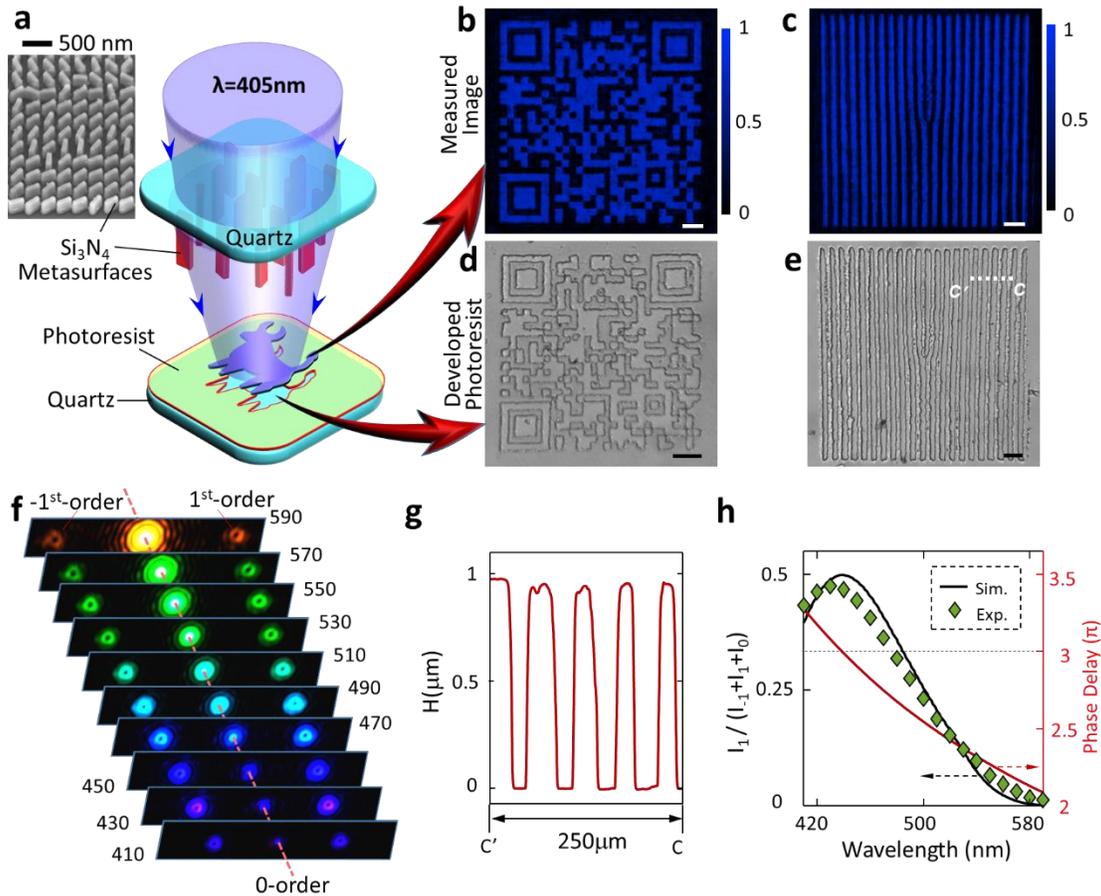

**Figure 4. Demonstration of computer-generated-holography lithography**. (**a**) Sketch for holographic lithography by using silicon nitride ($Si_3N_4$) metasurfaces with the SEM image shown in the insert. (**b-c**) Measured intensity profiles of a two-dimensional barcode (**b**) and a fork grating (**c**).

The design and characterization details are provided in Section 7 of Supplementary Materials. Scale bars: 10 μm. **(d-e)** Microscopy images of the developed photoresists with the barcode **(d)** and fork grating **(e)** patterns. The smallest feature sizes are ~25 μm in the patterned structures for both cases due to their different magnifications of ~7.3 (barcode) and ~10.3 (fork grating). Scale bars: 100 μm. **(f)** Diffraction patterns of the patterned fork grating at different wavelengths from 410 nm to 590 nm with an interval of 20 nm. Only three orders (0 and ±1) are shown here due to the limited detection region of our camera. **(g)** Measured height of the patterned photoresist along the dashed line $CC'$ in **(e)**. **(h)** Simulated (curve) and measured (green diamond) ratio of the 1-order diffracted intensity to the total intensity of three (0 and ±1) orders. The predicted phase delay between the patterned or unpatterned region in the developed fork grating is shown on the right axis.

# Supplementary Materials for

# Phase-probability shaping for speckle-free holographic lithography

*Dong Zhao[#], Weiwei Fu[#], Ziqin Li, Jun He, Kun Huang[*]*

Department of Optics and Optical Engineering, University of Science and Technology of China, Hefei, Anhui 230026, China

[#] *D. Z. and W. F.* contributed equally to this work.

[*]Corresponding authors: K. H. (**huangk17@ustc.edu.cn**)

## Table of Contents

**Section 1. Statistical properties of optical speckles generated by a random phase mask**

After high-coherence-laser illumination, a computer-generated hologram with a randomly distributed phase will yield an expected image. The speckles in the holographic image originate mainly from the random phase. Therefore, it is important to clarify the role of the random phase when investigating the statistical properties of the speckles. Note that, the random phase in the hologram has a special probability distribution from –π to π. In most cases, it is difficult to give an analytical description of the phase probability. To discuss the properties of the speckles conveniently, here we assume that the probability of phase encoded in the hologram has a Gaussian distribution, which exists in most random processes.

We begin by calculating the statistical properties of speckles that are generated via optical diffraction from a random pure-phase mask, as sketched in Fig. 1a. Assuming that the phase $\varphi_n$ at the position $(x_n, y_n)$ of the mask has the identical probability distribution with a Gaussian shape of $P(\varphi) = \frac{1}{\sqrt{2\pi}\sigma} e^{-\frac{\varphi^2}{2\sigma^2}}$ (where $\sigma$ denotes its standard deviation), one can obtain the electric field at the position $(x, y, z)$ of target plane[1]

$$E = \frac{-1}{2\pi}\sum_{n=1}^{N} e^{i\varphi_n}\left(ik - \frac{1}{R}\right)\frac{e^{ikR}}{R}\frac{z}{R}\Delta x \Delta y = \sum_{n=1}^{N} a_n e^{i(\varphi_n + \varphi_{0n})}, \quad (S1)$$

where $N$ is the number of pixels in the phase mask, the wave number $k=2\pi/\lambda$, $\lambda$ is the wavelength of light, $R^2=(x-x_n)^2+(y-y_n)^2+z^2$, $a_n=z\Delta x\Delta y(k^2R^2+1)^{1/2}/(2\pi R^3)$ and $\varphi_{0n}=-kR+\tan^{-1}(kR)$. Because each phase $\varphi_n$ is random with the Gaussian probability distribution $P(\varphi)$, the real ($u = \sum_{n=1}^{N} a_n cos(\varphi_n + \varphi_{0n})$) and imagery ($v = \sum_{n=1}^{N} a_n sin(\varphi_n + \varphi_{0n})$) parts are also the Gaussian random variables, implying that the electric field $E$ is a bivariate Gaussian distribution. According to the central limit theorem for a large $N$ (512×512 used in this work), the joint probability density function between $u$ and $v$ is expressed as[2]

$$P_{u,v}(u,v) = \frac{1}{2\pi\sigma_u\sigma_v\sqrt{1-\rho_{u,v}^2}} exp\left[-\frac{\left(\frac{u-\bar{u}}{\sigma_u}\right)^2 + \left(\frac{v-\bar{v}}{\sigma_v}\right)^2 - 2\rho_{u,v}\cdot\frac{u-\bar{u}}{\sigma_u}\cdot\frac{v-\bar{v}}{\sigma_v}}{2(1-\rho_{u,v}^2)}\right], \quad (S2)$$

where the average of the variable $X$ is defined as $\bar{X} = \int X \cdot P(\varphi)d\varphi$, $\sigma_X^2 = \overline{(X - \bar{X})^2}$ and $\rho_{u,v} = \overline{(u - \bar{u})(v - \bar{v})}/(\sigma_u\sigma_v)$. Eq. (S2) gives the probability correlating both the real and imagery parts of the field.

To derive the intensity probability from Eq. (S2), a coordinate transformation of $A = \sqrt{u^2 + v^2}$ and $\theta = tan^{-1}(v/u)$ is employed, yielding an $A$-$\theta$-joint probability $P_{A,\theta}(A, \theta) = P_{u,v}(u, v) \cdot |\mathbf{J}|$ (where the Jacobian determinant $\mathbf{J}$ in this transformation has a magnitude of $|\mathbf{J}| = A$)[3]. Note that, $A$ and $\theta$ are the amplitude and phase of the electric field $E$, respectively. Because the phase $\theta$ is not considered in the statistics of intensity, the integration with respect to $\theta$ results in an amplitude-dependent probability $P_A(A) = \int_{-\pi}^{\pi} P_{A,\theta}(A, \theta)d\theta$. Considering that the intensity $I=A^2$ is single valued for $A\geq 0$, a monotonic probability transformation from $A$ to $I$ yields the intensity probability at the position $(x, y, z)$ of interest

$$P_I(x,y,z,I) = P_A(\sqrt{I})\left|\frac{dA}{dI}\right| = \frac{1}{2}\int_{-\pi}^{\pi} P_{\sqrt{I}\cos\theta,\sqrt{I}\sin\theta}(\sqrt{I}\cos\theta,\sqrt{I}\sin\theta)d\theta, \quad (S3)$$

where $P_{\sqrt{I}\cos\theta,\sqrt{I}\sin\theta}$ is substituted directly by Eq. (S2).

In general, for a speckle, its intensity probability within a closed region of $\Sigma$ at a z-cut plane is used frequently in practice. Since two different positions in a speckle have the independent local probability of intensity, we can evaluate the total intensity probability in an interested region of the speckle by using

$$P_I(z,I) = \frac{1}{S}\iint_\Sigma P_I(x,y,z,I)dxdy$$

$$= \frac{1}{S}\iint_\Sigma \left\{\frac{1}{4\pi\sigma_u\sigma_v\sqrt{1-\rho_{u,v}^2}}\int_{-\pi}^{\pi}\exp\left[-\frac{\left(\frac{\sqrt{I}\cos\theta-\bar{u}}{\sigma_u}\right)^2 + \left(\frac{\sqrt{I}\sin\theta-\bar{v}}{\sigma_v}\right)^2 - 2\rho_{u,v}\frac{(\sqrt{I}\cos\theta-\bar{u})(\sqrt{I}\sin\theta-\bar{v})}{\sigma_u\sigma_v}}{2(1-\rho_{u,v}^2)}\right]d\theta\right\}dxdy, \quad (S4)$$

where $S$ is the area of the region. Eq. (S4) is one of the most important theoretical results because it can predict the intensity probability of the speckle in the region of interest. Due to the high complexity in the integral of Eq. (S4), the analytical solution is not available for most cases. Hence, the numerical solution of Eq. (S4) is employed in this work by carrying out the numerical integration in MATLAB software. In Eq. (S4), the parameters $\bar{u}$, $\bar{v}$, $\sigma_u$, $\sigma_u$ and $\rho_{u,v}$ are the functions of $x$, $y$ and $z$ but have no dependence on the variable $\theta$, so the numerical integration can be achieved with high accuracy. To avoid any confusion, we provide the detailed derivation of these parameters as follows:

$$\overline{\cos\varphi_n} = \int_{-\infty}^{\infty}\cos\varphi_n\frac{1}{\sqrt{2\pi}\sigma}e^{-\frac{\varphi_n^2}{2\sigma^2}}d\varphi_n = e^{-\frac{\sigma^2}{2}} \quad (S5)$$

$$\overline{\sin\varphi_n} = \int_{-\infty}^{\infty}\sin\varphi_n\frac{1}{\sqrt{2\pi}\sigma}e^{-\frac{\varphi_n^2}{2\sigma^2}}d\varphi_n = 0 \quad (S6)$$

$$\overline{\cos2\varphi_n} = \int_{-\infty}^{\infty}\cos2\varphi_n\frac{1}{\sqrt{2\pi}\sigma}e^{-\frac{\varphi_n^2}{2\sigma^2}}d\varphi_n = e^{-2\sigma^2} \quad (S7)$$

$$\overline{\sin2\varphi_n} = \int_{-\infty}^{\infty}\sin2\varphi_n\frac{1}{\sqrt{2\pi}\sigma}e^{-\frac{\varphi_n^2}{2\sigma^2}}d\varphi_n = 0 \quad (S8)$$

$$\bar{u} = \sum_{n=1}^{N}a_n\overline{\cos(\varphi_n+\varphi_{0n})} = \sum_{n=1}^{N}(a_n\cos\varphi_{0n}\overline{\cos\varphi_n} - a_n\sin\varphi_{0n}\overline{\sin\varphi_n}) = \sum_{n=1}^{N}(a_n\cos\varphi_{0n}\overline{\cos\varphi_n}) \quad (S9)$$

$$\bar{v} = \sum_{n=1}^{N}a_n\overline{\sin(\varphi_n+\varphi_{0n})} = \sum_{n=1}^{N}(a_n\sin\varphi_{0n}\overline{\cos\varphi_n} + a_n\cos\varphi_{0n}\overline{\sin\varphi_n}) = \sum_{n=1}^{N}(a_n\sin\varphi_{0n}\overline{\cos\varphi_n}) \quad (S10)$$

$$\overline{u^2} = \overline{\left[\sum_{n=1}^{N}a_n\cos(\varphi_n+\varphi_{0n})\right]\cdot\left[\sum_{m=1}^{N}a_m\cos(\varphi_m+\varphi_{0m})\right]}$$

$$= \overline{\sum_{n=1,m=n}^{N}a_n\cos(\varphi_n+\varphi_{0n})a_m\cos(\varphi_m+\varphi_{0m})} + \overline{\sum_{n=1}^{N}\sum_{m=1,m\neq n}^{N}a_m\cos(\varphi_m+\varphi_{0m})a_n\cos(\varphi_n+\varphi_{0n})}$$

$$= \overline{\sum_{n=1}^{N}a_n^2\cos^2(\varphi_n+\varphi_{0n})} + \overline{\sum_{n=1}^{N}\sum_{m=1,m\neq n}^{N}a_ma_n(\cos\varphi_m\cos\varphi_{0m}-\sin\varphi_m\sin\varphi_{0m})(\cos\varphi_n\cos\varphi_{0n}-\sin\varphi_n\sin\varphi_{0n})}$$

$$= \sum_{n=1}^{N} \frac{a_n^2}{2} \left[\overline{\cos(2\varphi_n + 2\varphi_{0n})} + 1\right] + \sum_{n=1}^{N} \sum_{m=1,m\neq n}^{N} a_m a_n (\overline{\cos\varphi_m} \cos\varphi_{0m})(\overline{\cos\varphi_n} \cos\varphi_{0n})$$

$$= \sum_{n=1}^{N} \frac{a_n^2}{2} \left[\overline{\cos(2\varphi_n)} \cos(2\varphi_{0n}) + \overline{\sin(2\varphi_n)} \sin(2\varphi_{0n}) + 1\right] + \sum_{n=1}^{N} \sum_{m=1,m\neq n}^{N} a_m a_n (\overline{\cos\varphi_m} \cos\varphi_{0m})(\overline{\cos\varphi_n} \cos\varphi_{0n})$$

$$= \sum_{n=1}^{N} \frac{a_n^2}{2} \left[\overline{\cos(2\varphi_n)} \cos(2\varphi_{0n}) + 1\right] + \left[\sum_{m=1}^{N} a_m (\overline{\cos\varphi_m} \cos\varphi_{0m}) \sum_{n=1}^{N} a_n (\overline{\cos\varphi_n} \cos\varphi_{0n}) - \sum_{n=1}^{N} a_n^2 (\overline{\cos\varphi_n} \cos\varphi_{0n})^2\right]$$

$$= \left\{\sum_{n=1}^{N} \frac{a_n^2}{2} \left[\overline{\cos(2\varphi_n)} \cos(2\varphi_{0n}) + 1\right]\right\} + \overline{\cos\varphi_n}^2 \left\{\left[\sum_{n=1}^{N} a_n \cos\varphi_{0n}\right]^2 - \sum_{n=1}^{N} (a_n \cos\varphi_{0n})^2\right\} \quad (S11)$$

$$\overline{v^2} = \overline{\left[\sum_{n=1}^{N} a_n \sin(\varphi_n + \varphi_{0n})\right] \cdot \left[\sum_{m=1}^{N} a_m \sin(\varphi_m + \varphi_{0m})\right]}$$

$$= \left\{\sum_{n=1}^{N} \frac{a_n^2}{2} \left[1 - \overline{\cos(2\varphi_n)} \cos(2\varphi_{0n})\right]\right\} + \overline{\cos\varphi_n}^2 \left\{\left[\sum_{n=1}^{N} a_n \sin\varphi_{0n}\right]^2 - \sum_{n=1}^{N} (a_n \sin\varphi_{0n})^2\right\} \quad (S12)$$

$$\overline{uv} = \overline{\left[\sum_{n=1}^{N} a_n \cos(\varphi_n + \varphi_{0n})\right] \cdot \left[\sum_{m=1}^{N} a_m \sin(\varphi_m + \varphi_{0m})\right]}$$

$$= \overline{\cos 2\varphi_n} \left\{\sum_{n=1}^{N} \frac{a_n^2}{2} \sin 2\varphi_{0n}\right\} + \overline{\cos\varphi_n}^2 \left\{\left[\sum_{n=1}^{N} a_n \cos\varphi_{0n}\right] \cdot \left[\sum_{m=1}^{N} a_m \sin\varphi_{0m}\right] - \sum_{n=1}^{N} (a_n^2 \sin\varphi_{0n} \cos\varphi_{0n})\right\} \quad (S13)$$

$$\sigma_u^2 = \overline{(u - \bar{u})^2} = \overline{u^2} - \bar{u}^2 \quad (S14)$$

$$\sigma_v^2 = \overline{(v - \bar{v})^2} = \overline{v^2} - \bar{v}^2 \quad (S15)$$

$$\rho_{u,v} = \overline{(u - \bar{u})(v - \bar{v})}/(\sigma_u \sigma_v) = (\overline{uv} - \bar{u} \cdot \bar{v})/\sqrt{(\overline{u^2} - \bar{u}^2)(\overline{v^2} - \bar{v}^2)} \quad (S16)$$

Now, we point out several issues when applying the Eq (S4) (*i.e.*, Eq. (1) in the main text) in solving the practical problems. Firstly, the equation is valid only when each scattering point in the mask has the same Gaussian probability distribution. Other probability distributions beyond the Gaussian shape should be divided into different Gaussian probability distributions with weighted coefficients. Thus, one can use the multivariate mathematical analysis[4] to calculate the joint probability density. Note that, all the mathematical formulas are based on the condition that all the positions in the phase mask have the same probability density (*i.e.*, Gaussian or non-Gaussian distribution). For the designed phase in this work, the phase probability density varies in the entire phase mask so that one cannot use the Eq. (S4) directly to calculate its statistical analysis. However, we must emphasize that the basic conclusions (*e.g.*, the uniform probability distribution increases the speckles and narrowly distributed probability decreases the speckle contrast) obtained by using Eq. (S4) are valid for most holograms, including Fresnel and Fraunhofer holograms that have different imaging distances of *z*. The statistical properties in the holographic images are calculated by using the diffraction patterns predicted by the Rayleigh-Sommerfeld integrals[1].

Secondly, Eq. (S4) can yield the phase probability density for some special cases, such as $\sigma=0$ and $\sigma=\infty$. When $\sigma=0$, it results in $\sigma_u=\sigma_v=\rho_{u,v}=0$. In this case, Eq. (S4) becomes the delta function, which means that the intensity is a constant with great uniformity. This is related to the case where the aperture of the mask is infinitely large. In fact, for a finite-size mask, the diffraction pattern from the edge of the mask leads to the fluctuation of intensity. Therefore, Eq. (S4) has not considered the weak diffraction of the boundary, which dominates the intensity for small $\sigma$. When $\sigma=\infty$, the resulting parameters are $\sigma_u^2 = \sigma_v^2 = \sum_{n=1}^{N} a_n^2/2$ and $\rho_{u,v}=0$. Thus, Eq. (S4) is simplified into $P_I(z, I) = \frac{1}{S} \iint_\Sigma e^{-I/(2\sigma_u^2)}/(2\sigma_u^2)\, dxdy$, where $\sigma_u$ is dependent on the position at the cut plane of interest. However, when the speckle region of interest is not large, $\sigma_u$ has no significant variation so it can be taken as a constant. Thus, the final intensity probability density $P_I(z, I)$ obeys the Rayleigh distribution[5], which is a key feature of a fully developed speckle. It means that the uniformity of the speckle is poor for a large $\sigma$.

Thirdly, Eq. (S4) only provides the probability density of intensity in the speckle pattern. In fact, in many applications, one can also obtain the probability density of the phase in the speckle pattern by integrating $P_{A,\theta}(A, \theta)$ with respect to $A$ and implementing some similar derivations shown above. Thus, we can easily evaluate the statistical properties of the speckle at any position along the propagating direction of light.

## Section 2. Reconstructing the speckle-free holograms by using our MAPS algorithm

To show more details about the designed results by our MAPS algorithm, we provide the efficiency, RMSE and standard deviation (SD) of the simulated images after 1500 iterations in Supplementary Figs. 1a-1c. Referring to Fig. 2d in the main text, we can find that, the uniformity (evaluated by RMSE and SD) is good in the singularity-free region, while its efficiency is slightly lower than those in the singularity region.

Now, we reveal the fundamental origin of the different phase probability distributions for the singularity-free and singularity cases. As shown in Supplementary Figs. 1d and 1e, the designed phase profiles for the singularity-free (A) and singularity (B) cases show blurred contours of the expected holographic image, which is caused by the fact that the Fresnel hologram is demonstrated here. Within the region of the image contour, the phase has nearly the same phase probability distribution, implying that the direct projection of the phase with the image contour leads to the expected image. Such a feature also exists in the phase (see Fig. 3b in the main text) designed by using the GS algorithm. However, their differences come from the region outside the image contour. The phase in the GS case (Fig. 3b in the main text) is uniformly distributed from $-\pi$ to $\pi$ outside the image contour, so that light shining outside the image contour is convergent to the holographic image and then interferes with the diffraction light inside the image contour. Thus, such interference leads to the creation of phase singularities due to the irregular shape of the target image for the GS case and singularity (MAPS) case. In comparison, for the singularity-free case designed via our MAPS algorithm, the phase outside the image contour is valued locally with a narrow probability density, which avoids the convergence of light onto the holographic image. Instead, the narrowly distributed phase (without introducing fully developed speckles that occur for the uniform-distribution phase, as predicted in Eq. (1) and Eq. (S4)) outside the image contour will diffract light away from the image region at the target plane, which is the fundamental reason for the low efficiency in the speckle-free holograms. This implies that the diffraction efficiency of the speckle-free hologram is approximated by the ratio (26% in this design of the 'bull' pattern) of the image size to the mask size, which is comparable to the simulated efficiency of ~30% (see Supplementary Fig. 1a). The above analysis offers an insightful understanding of reconstructing speckle-free holographic images.

Based on this recognition, we predict that the phase mask must have the larger size than the reconstructed speckle-free image. The proofs are twofold. The first one comes from the Fraunhofer holograms[6], where the size of the image is much larger than the size of the phase mask. Moreover, in many reported works including traditional multi-level diffractive optical elements[7] and recently reported metasurfaces[8,9], the designed phase in the Fraunhofer holograms is uniformly distributed from $-\pi$ to $\pi$, which will definitely lead to the speckles according to our theoretical prediction by Eq. (1) in the main text or Eq. (S4) in the Supplementary Materials. To the best of our knowledge, we have not found Fraunhofer holograms that exhibit speckle-free feature under high-coherence illumination. The second proof is observed by using our design example, where the image is slightly larger than the phase mask for a Fresnel hologram. It is designed by using our MAPS algorithm with $w_{SD}=1$ and $w_{Eff}=0$. Supplementary Figs. 1f and 1g show the designed phase and its relative holographic image, respectively. In the simulated image, one can clearly see the phase singularities outside the dashed square (denoting the phase mask) but the high-uniformity intensity inside, thereby suggesting the second evidence.

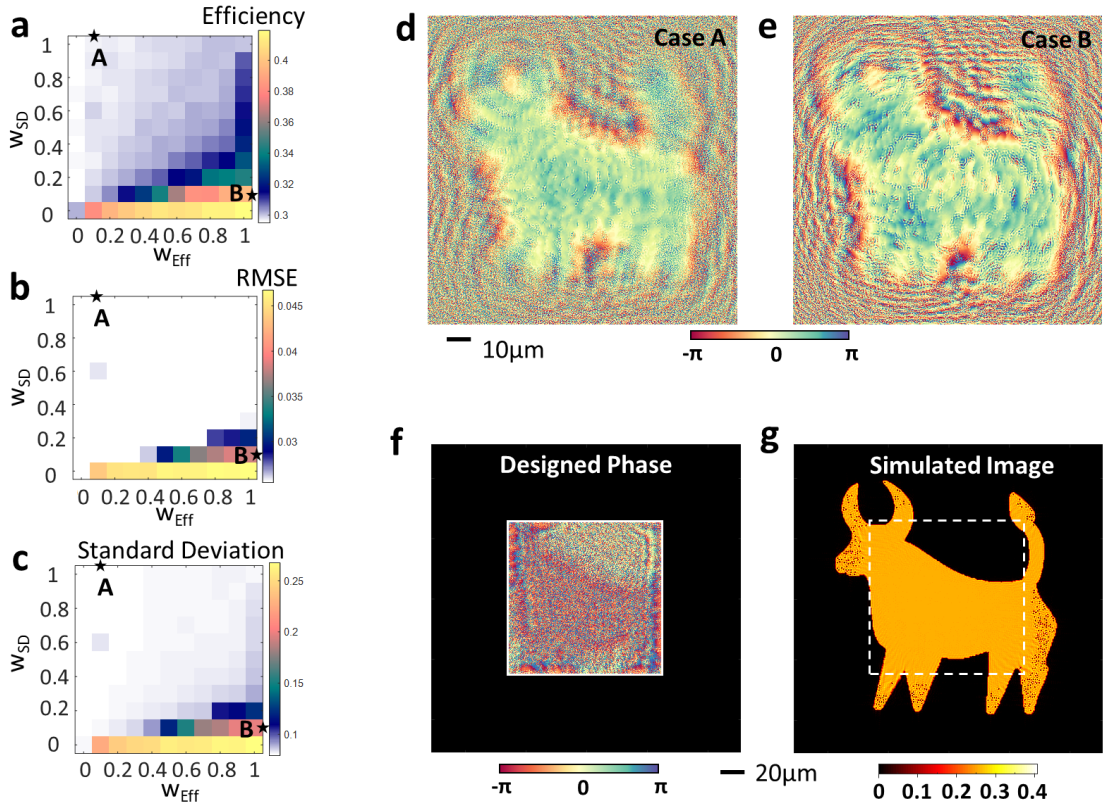

**Supplementary Figure 1 | Design results of the hologram by using the MAPS algorithm. (a-c)** Simulated efficiency **(a)**, RMSE **(b)** and standard deviation **(c)** of the designed holograms by scanning $w_{SD}$ and $w_{Eff}$ from 0 to 1 in our MAPS algorithm. **(d-e)** Simulated phase profiles for the singularity-free **(d)** and singularity **(e)** cases as discussed in Fig. 2 of the main text. **(f-g)** Designed phase **(f)** and its relative simulated image **(g)** at the target

plane when the size of the image is larger than that of the phase mask. In our simulations, we have a pixel pitch of 250 nm×250 nm, the imaging distance of $z$=250 μm, and a mask size of 128 μm×128 μm. The dashed square denotes the boundary of the phase mask, which is smaller than the target image.

**Section 3. Design of geometric metasurfaces**

To realize the designed phase, we employ geometric metasurfaces composed of crystalline silicon nanobricks on a sapphire substrate, as sketched in Supplementary Fig. 2a. In such metasurfaces, the silicon nanobricks work as a miniaturized half-waveplate[10,11] that can convert the incident circular polarization into its crossed one with an additional geometric phase of twice the in-plane rotating angle $\theta$. Such a geometric phase is used to realize the designed phase, which is divided into 128 levels with an interval of $2\pi/128$ between two neighboring levels. Thus, such a phase modulation with high accuracy is helpful in reducing the experimental errors[12]. To maximize the conversion efficiency, we simulate the optical performance of the nanobricks by using the finite-difference time-domain (FDTD) method. In our FDTD model, the metasurfaces have a pixel pitch of 250 nm×250 nm with the periodic boundaries along $x$ and $y$ directions while the perfect-matching layers (PMLs) are employed to eliminate the reflection from the top and bottom materials. Considering the low absorption of crystalline silicon at visible wavelengths, we choose an operating wavelength of 633 nm during the optimization of conversion efficiency. The thickness of such c-Si film on the sapphire substrate is 300 nm, which is fixed in our simulation model. In this case, we can only optimize the length and width of the nanobricks to maximize the conversion efficiency. By scanning its width and length, we obtain the simulated conversion efficiency as shown in Supplementary Fig. 2b, which indicates an efficiency of >90% for the 100-nm width and the 150-nm length (as denoted by the asterisk) for the fabrication. To evaluate the possible errors, we also simulate the conversion efficiency when the rotating angle $\theta$ changes from 0 to 180 degrees. The simulated results in Supplementary Fig. 2c reveal a sine-like shape with a fluctuation of ~0.02, which has no significant influence on the holographic image. To show it numerically, we simulate the diffraction patterns without (Supplementary Fig. 2d) or with (Supplementary Fig. 2e) such a fluctuation. The calculated RMSEs and speckle contrasts for both cases are nearly identical. Therefore, we confirm that the error caused by the conversion efficiency will not destroy the uniformity of the holographic images. For all the simulations throughout this work, such efficiency error is not considered.

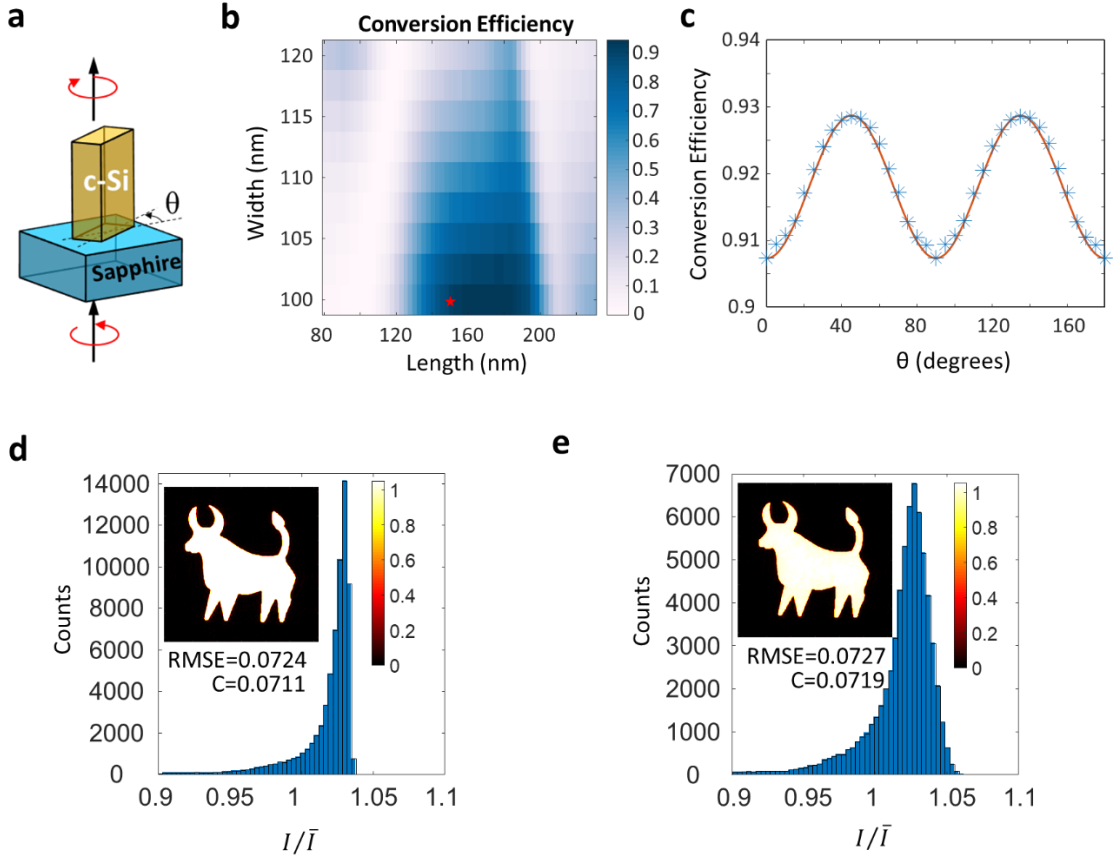

**Supplementary Figure 2 | Design details of silicon-based geometric metasurfaces working in the transmission mode.** (a) Sketch of a nanobrick in a unit cell for this geometric metasurface. (b) Simulated conversion efficiency of the nanobricks with different widths and lengths at a fixed height of 300 nm. (c) Dependence of the conversion efficiency on the rotation angle $\theta$ (as sketched in (a)) of the nanobricks. A sine-shape fitting (curve) of the simulated efficiency (asterisks) is provided here for a better evaluation of the variation in the conversion efficiency. (d-e) Intensity probability density (bars) of the simulated diffraction patterns (inserts) without (d) and with (e) considering the error from the conversion efficiency. Although the error caused by the varied conversion efficiency leads to the slight extension of the intensity-probability density, their RMSEs and speckle contrasts (C) are nearly the same, as denoted in the figures.

### Section 4. Fabrication of silicon metasurfaces

The metasurfaces are fabricated based on the process of deposition, patterning, lift off, and etching. The sapphire substrate is first deposited with a 300 nm-thick single-crystal silicon film, which determines the height of the designed Si nanorods. After a 100 nm-thick positive electron-beam resist (AR-P 6200) is coated and baked on the silicon film, the desired structures are patterned using electron beam lithography (JEOL, JBX 6300FS) with an accelerating voltage of 100 kV. After development, a 10 nm chromium layer is deposited on the sample using an E-beam evaporator (Kurt J. Lesker, PVD75 Proline) followed by a lift-off procedure. Then, the 300 nm-thick Si layer without chromium hard mask is etched by an inductively coupled plasma-reactive ion etching (ICP-RIE) system (Oxford, Plasma Pro System100 ICP380). The residual chromium mask is removed by chromium etchant at last. The SEM images of the fabricated samples are shown in Figs. 3f, 3g and 4a of the main text. The good profiles and clear edges are obtained, implying the good fabrication. In addition, for the silicon nitride metasurfaces, their fabrication details are the same with the above processes.

## Section 5.  Experimental characterization of fabricated silicon metasurfaces

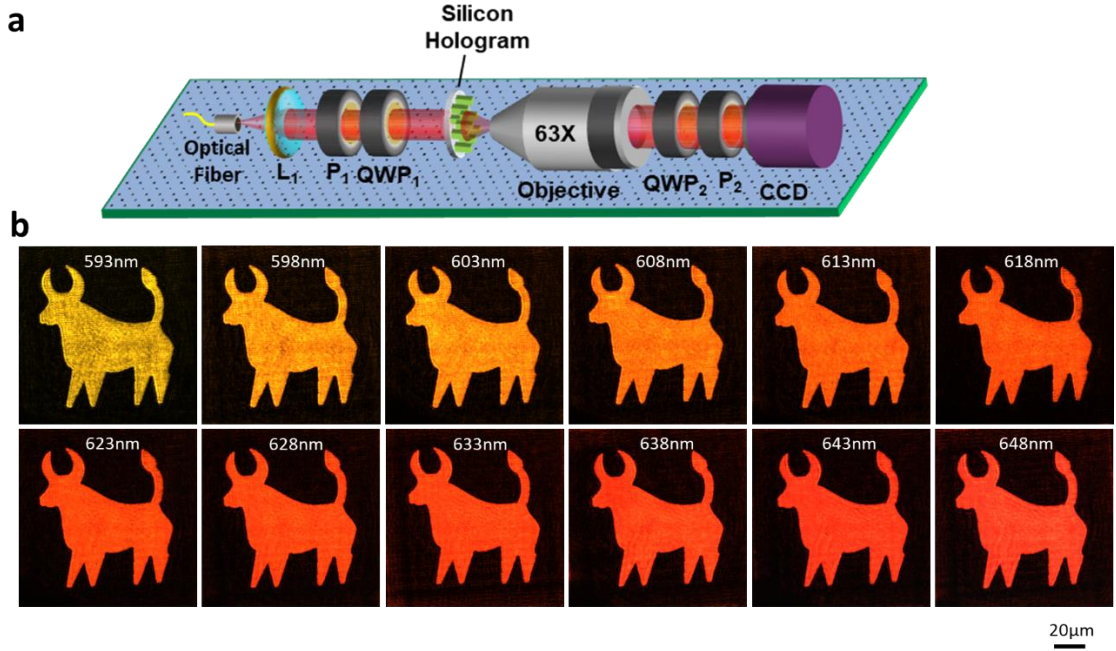

**Supplementary Figure 3 | Experimental setup (a) and the relative measured images (b) at the different wavelengths.**

To test the optical performance of the fabricated silicon metasurfaces, we have built an experimental setup, as shown in Supplementary Fig. 3a. A supercontinuum laser (Superk Fianium) filtered by an acousto-optic modulator (with a bandwidth of 2.5 nm-5 nm at the wavelengths of interest, SuperK Select) is used to illuminate the fabricated samples after collimated by a lens ($L_1$). A linear polarizer ($P_1$) and a quarter waveplate ($QWP_1$) are used to obtain the required circular polarization (CP). After passing through the metasuface, the CP light is collected by using an objective lens (Zeiss, 0.95NA 63X), which then projects the generated holographic image onto a charge-coupled-device camera (LBtek, STC-MCS500POE) after a CP analyzer (composed of a quarter waveplate $QWP_2$ and a linear polarizer $P_2$). By changing the filtered wavelengths, we can investigate the broadband response of our proposed holograms. Supplementary Fig. 3b shows the raw data of the measured images at the different wavelengths. Their detailed data analyses can be found in Fig. 3 of the main text. Based on all these experimental data, we confirm the validity of the proposed holograms.

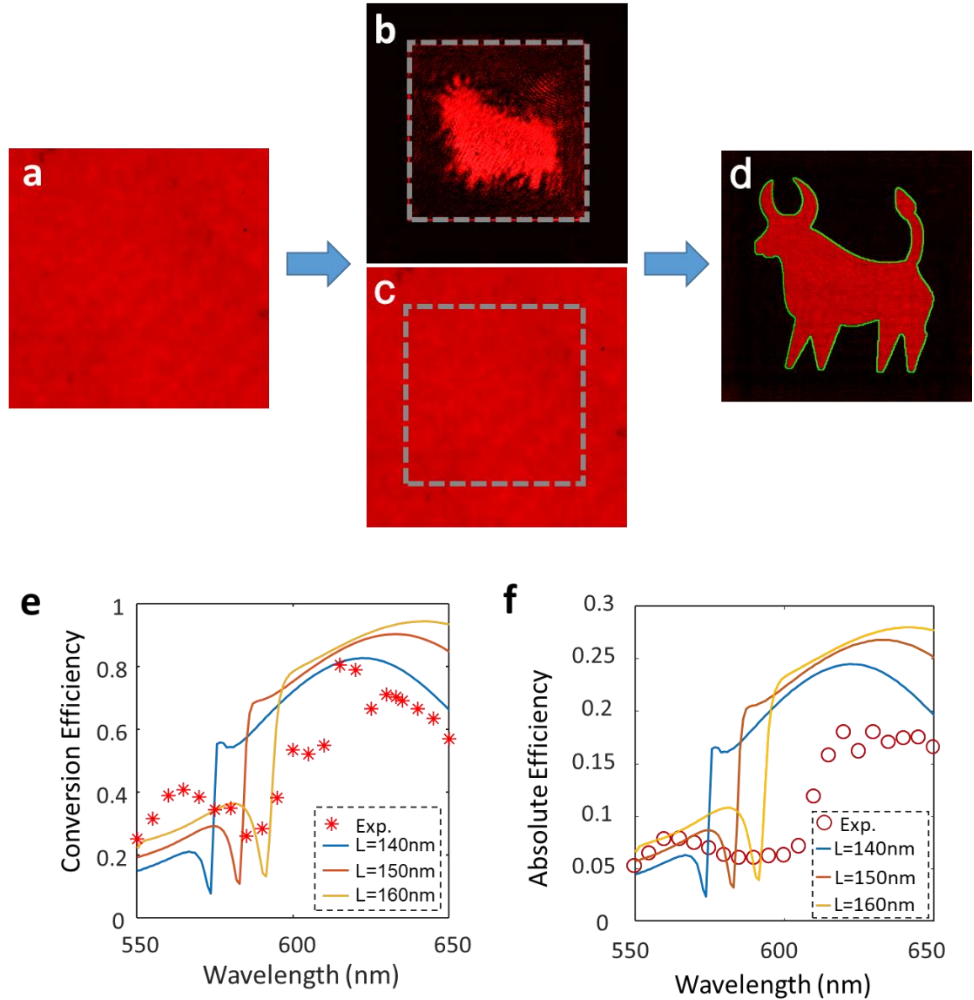

**Supplementary Figure 4 | Efficiency characterization of the demonstrated holograms. (a-d)** Flow chart of our efficiency characterization. **(e)** Simulated (curves) and measured (asterisks) conversion efficiency at the wavelengths of interest. The simulated conversion efficiency profiles of the nanobricks with different lengths are also provided for a better comparison because the fabricated nanobricks might have a slight deviation from the designed dimension of L=150 nm. **(f)** Simulated (curves) and measured (asterisks) absolute efficiency of the fabricated holograms. The absolute efficiency is defined as the ratio of the power of the achieved holographic image to the total power of the transmitted through a bare (without any nanostructure) substrate.

The optical efficiency of the metasurfaces has also been characterized by using the optical setup in Supplementary Fig. 3a. To measure its conversion efficiency, we have fabricated an array of nanobricks. After putting it into the measurement system, we firstly record the intensity on the bare quartz (beyond the region of the nanobrick array) by using the CCD camera, where the CP analyzer (composed of $QWP_2$ and $P_2$) is tuned to maximize the intensity of the CCD (i.e., the incident light has the same polarization with the transmitted light), see Supplementary Fig. 4a. Then, the nanobrick array is moved into the field of view. Tuning the CP analyzer, we make the light beyond the nanobrick array the darkest (see Supplementary Fig. 4b), which means that the transmitted light through the nanobrick array has orthogonal polarization to the incident one. Thus, the total intensity located within the area (*e.g.*, the dashed square in Supplementary Fig. 4b) of the array is the converted intensity, which is labeled by $I_{out}$. Correspondingly, the total intensity located in the same area of the dashed square in Supplementary Fig. 4a is its relative incident intensity $I_{in}$. Therefore, the experimental conversion efficiency is calculated as $I_{out}/I_{in}$, which is shown in Supplementary Fig. 4e. Compared with the simulated conversion efficiency, the experimental data is lower, which might be caused by the imperfect fabrication (such as small particles created during the etching process) and the mismatch between the simulated and practical refractive indices. Nevertheless, the measured conversion efficiency at the wavelengths from 608 nm to 648 nm is larger than 50%, thus leading to a weak background and better uniformity.

To measure the absolute efficiency of the metasurface holograms, we use the hologram sample (as shown in Supplementary Figs. 4a-4d) instead. By keeping the completely identical steps as discussed above, we continue to measure the intensity of the holographic image by projecting the created images onto the CCD, as shown in Supplementary Fig. 4d. Only the intensity encircled within the image (*i.e.,* the bull pattern) is taken as the final intensity of the output, which is labelled by $I_{out}$. Thus, the absolute efficiency is defined as the ratio of $I_{out}/I_{in}$, which is shown in Supplementary Fig. 4f. The measured absolute efficiency for $\lambda>608$ nm is ~17%, which is lower than the simulated efficiency that can be calculated by the product of the conversion efficiency and the diffraction efficiency of the hologram. To further enhance the efficiency, one can use less-absorption materials (such as $TiO_2$, SiN, GaP, etc) at the red-light wavelengths as well as good fabrication of the nanobricks. In despite of the low efficiency in our proposed metasurface holograms, the holographic images have much higher intensity than that of the scattering light, so that the subsequent exposure in the following lithography process is less influenced, as confirmed by our demonstrated lithography in the main text.

**Section 6. Experimental factors influencing the uniformity of the holographic images**

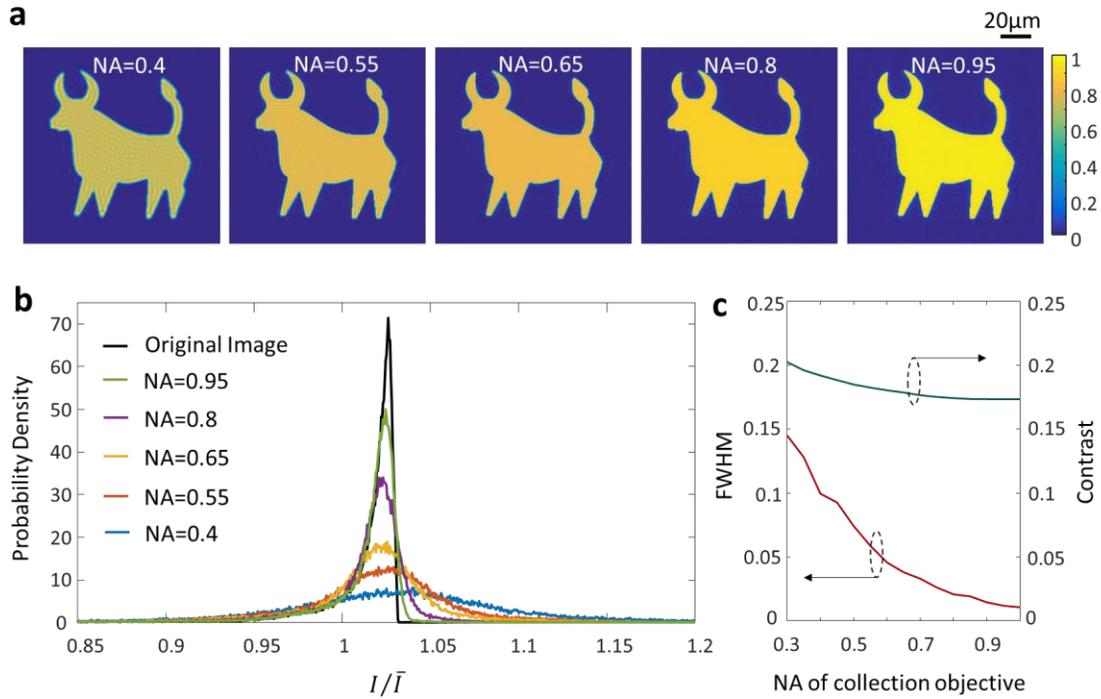

**Supplementary Figure 5 | Simulations of the uniformity of the holographic images collected by different objective lenses. (a)** Simulated images at the different NAs. **(b)** Simulated probability density of intensity. **(c)** Calculated FWHMs (left) of the probability density and the speckle contrasts (right) of the simulated images.

Due to the importance of intensity uniformity in practical applications such as lithography, we provide a detailed discussion of the possible factors in the experiment. Firstly, the large bandwidth of the illuminating light is important to increase the uniformity but blur the edge. In this work, the bandwidth of the lasers filtered by the acoustic-optic modulator is 2.5 nm – 5 nm, which has not blurred the edge significantly, as shown in Fig. 3o of the main text. Secondly, the alignment of the entire optical system, including the illumination and collection parts, is critical in achieving high uniformity in the experiment. The misalignment along any direction (*i.e.*, *x*, *y* and *z*) will lead to the fluctuation of intensity in the image. To eliminate the spatial misalignment, we have mounted the key elements onto the high-precision three-dimensional stages (*e.g.*, Newport linear transport stages). Third, the fabrication issue of the nanobricks is also the key factor in determining the uniformity. In the laboratory environment where all the fabrication tools are shared with the parameters changed frequently by various users, it is challenging to obtain stable devices. Actually, industrial tools such as deep-ultraviolet lithography are able to enhance the fabrication quality for better uniformity. Finally, the numerical apertures of the collection objectives are also important to maintain good uniformity. Despite their important role, the numerical apertures of the collection objectives are less considered in real experiments. To highlight it in more detail, we provide some simulations of the images (see Supplementary Fig. 5a) taken by the objectives with different numerical apertures. The simulations are implemented by using Fourier filtering, where light beyond the largest spatial frequencies for the numerical aperture of the objective is removed. One can find that the objectives with a larger NA will lead to better uniformity although all the inputs are the same. The probability density functions of the intensity in the filtered images shown in Supplementary Fig. 5b doubly confirm the above conclusion with the quantitative proofs because the probability density of intensity is narrow for a larger-NA objective. The calculated FWHM of the intensity probability density and the simulated speckle contrasts also suggest that a higher-NA objective should be used to avoid damage to the image uniformity. Therefore, in our experimental measures, a 0.95-NA objective lens is used to capture the holographic images, which have good uniformity, as shown in Supplementary Fig. 5a.

**Section 7. Demonstration of computer-generated-hologram lithography**

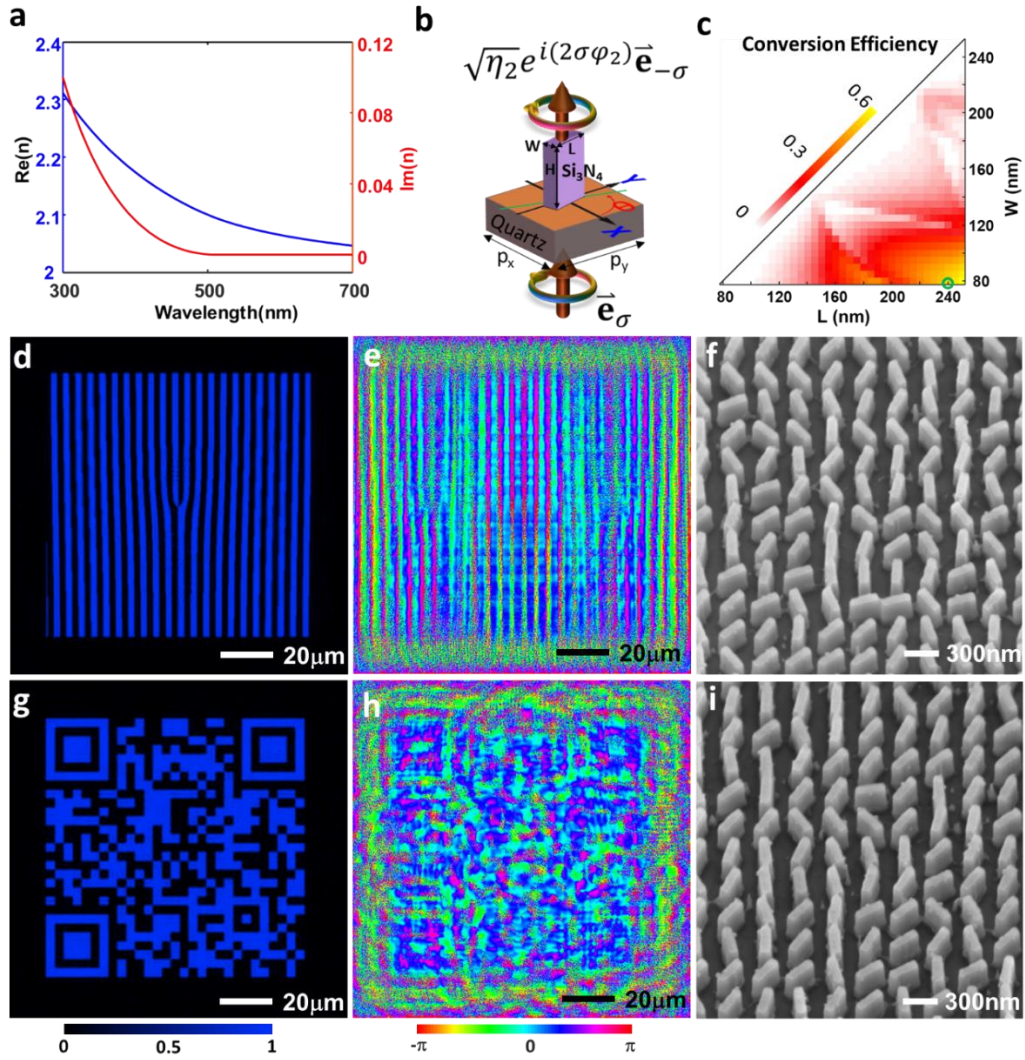

**Supplementary Figure 6 | Design details of holographic images for optical lithography. (a)** Measured refractive indices (with the real and imaginary parts shown on the left and right axis, respectively) of the prepared silicon nitride film on the quartz substrate. **(b)** Sketch of one unit cell in the proposed silicon nitride geometric metasurfaces that can convert one circular polarization into its cross one with an additional phase modulation of twice the rotation angle of the nanobricks. **(c)** The simulated conversion efficiency by scanning the widths and the lengths of the nanobricks at a fixed thickness of 330 nm. It is simulated by using the FDTD method. **(d-f)** Simulated intensity **(d)** of the holographic image (fork grating) by using its relative phase **(e)** that is designed via our MAPS algorithm. The SEM image of the relative fabricated sample is shown in **(f)**. **(g-i)** Simulated intensity **(g)** of the holographic barcode image reconstructed by using the phase profile in **(h)**. The geometric metasurfaces with the SEM image shown in **(i)** are used to realize the designed phase.

In our demonstration of holographic lithography, we use a working wavelength of 405 nm for better exposure of the photoresist. To enhance the optical efficiency of polarization conversion, silicon nitride with less absorption at λ=405 nm is employed as the material of the nanobricks in the geometric metasurfaces. Our prepared silicon nitride film (with an optimized thickness of 330 nm) deposited on a quartz substrate has measured refractive indices at wavelengths ranging from 300 nm to 700 nm, which indicates the refractive index of 2.169+0.01755$i$ at the wavelength of 405 nm. Although the imaginary part of the refractive index will lead to optical absorption, silicon nitride exhibits much better efficiency than high-loss silicon at this near-ultraviolet wavelength. So, the configuration of geometric metasurfaces with a pixel pitch of 260 nm×260 nm (see Supplementary Fig. 6b)allows us to optimize the optical conversion efficiency as high as 60% at the selected dimensions of $L$=240 nm and $W$=80 nm, as shown in Supplementary Fig. 6c. By using e-beam lithography with a dry etching process (as introduced above), we fabricated the samples and tested its experimental conversion efficiency of 30%, which is lower than the expected 60%. The main reason is the imperfect fabrication due to the residual nanoparticles around the patterned nanobricks.

The hologram is designed by using our MAPS algorithm with the parameters $w_{SD}$=1 and $w_{Eff}$=0.1. It is implemented with an iteration of 3000. To demonstrate the capability of holographic lithography, we design two patterns: a fork grating as a prototype of an optical device, and a randomly distributed two-dimensional barcode as a complicated pattern. By using these images, we can evaluate the ability of holographic lithography in patterning optical devices and complex structures, which are the key features in the lithography technique. The designed fork grating with a period of 4.5 μm (Supplementary Fig. 6d) is reconstructed at $z$=390 μm by using the relative phase in Supplementary Fig. 6e. From the designed phase, we can find that the phase jump of π is observed, which is helpful to create the sharp edge at the bright fringes. In our experimental demonstration, the phase level is also 128, which increases the accuracy of the holographic reconstruction. Supplementary Fig. 6f shows the SEM image of its related geometric metasurfaces. Similarly, the design and fabrication details of the barcode case are shown in Supplementary Figs. 6g-6i, where all the other parameters are the same except for the image. All the measured images are shown in Figs. 4b and 4c in the main text, which are in good agreement with the simulated results. Considering the demonstrated conversion efficiency of ~30% and the diffraction efficiency (~30%) of our designed holograms, the absolute efficiency is approximately 9% in the holographic reconstruction.

To demonstrate this for lithographic usage, we have built up the measurement system shown in Supplementary Fig. 7a. The 405-nm-wavelength light emitted from a semiconductor laser is reshaped into a quasi-Gaussian beam by using a spatial filtering that is composed of two confocal lenses and a pinhole (PH). A circular polarizer composed of a linear polarizer ($P_1$) and a quarter waveplate ($QWP_1$) is employed as the incident light of our fabricated silicon-nitride metasurface holograms. Due to the limited imaging distance of $z$=390 nm, we use a couple of objectives ($NA$=0.85, 60X, ZEISS; NA=0.3, 20X, ZEISS) to project the reconstructed images onto the photoresist (Microposit S1813, Shipley) that is spin-coated with a thickness of 950 nm on a quartz substrate. The collection objective lens with an NA of 0.85 is used to avoid damage to the uniformity in the image during the projection process, as discussed in Section 6 of Supplementary Materials. To remove the undesired background, we use another circular analyzer (composed of a $QWP_2$ and a linear polarizer $P_2$), which can theoretically be removed if the conversion efficiency is high. After considering the incident power on the metasurface hologram, the optical efficiency of metasurfaces, the transmission of both objective lenses, and the transmission of the $QWP_2$ and the $P_2$, we estimate that the power of the holographic image for the exposure is at the level of tens of microwatts.

To find the best exposure time, we expose the photoresist with different times and show the developed patterns in Supplementary Fig. 7b, which suggests a suitable exposure time of 10 seconds at the special incident power. It is worth nothing that the best exposure time will change if the power of the holographic image is different. So, it is quite important to find suitable exposure parameters based on optical systems, including stable sources and well-mounted elements. From the developed patterns in Supplementary Fig. 7b, we can conclude that, the insufficient exposure with less interacting time will lead to residual photoresist after the development process, while the over-exposure usually results in the fact that the photoresist beyond the desired patterns will be exposed to destroy the quality of the patterning process.

Furthermore, to make the right alignment between the holographic image and the photoresist plane, we use a high-precision 3-dimensional stage to tune the position of the photoresist. The initial position of the photoresist sample is roughly evaluated by using a CCD camera, so that we can find the right imaging position easily without a long-time search. In fact, such an alignment between the holographic images and the photoresist can be controlled by an automatic opto-mechatronic system in commercial lithographic systems.

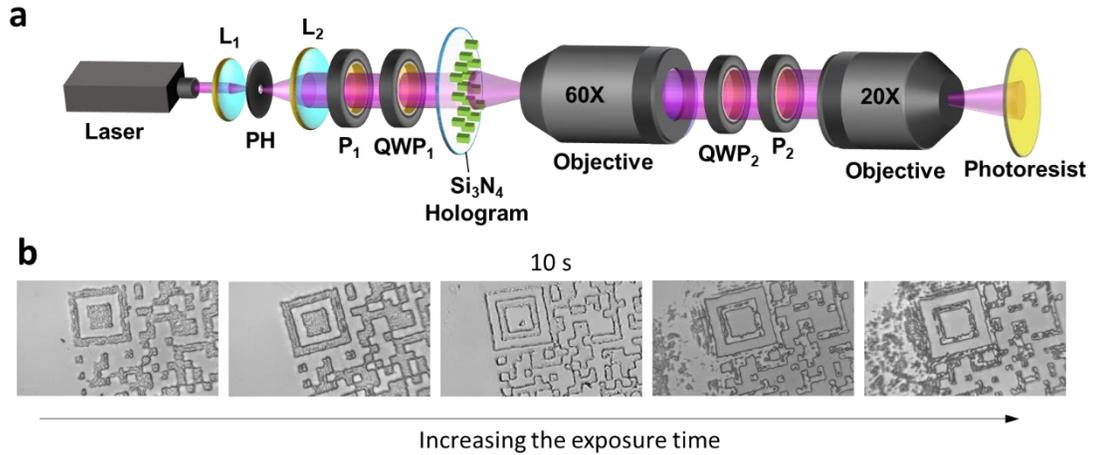

**Supplementary Figure 7 | Experimental results of holographic lithography. (a)** Optical experimental setup for characterizing the holographic lithography. PH: Pinhole; P: Polarizer; QWP: Quarter waveplate; **(b)** Microscopic images of the developed photoresist patterns under different exposure times.

## Section 8. Characterizing optical performance of the patterned devices

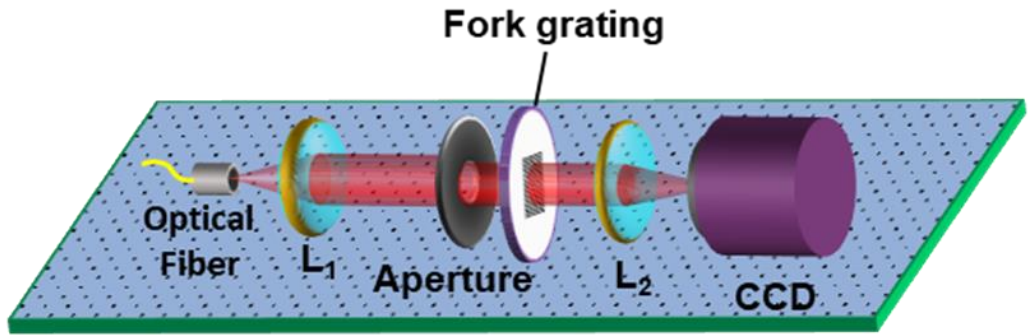

**Supplementary Figure 8 | Optical setup for characterizing the performance of the fork grating patterned by using our holographic lithography.**

    To investigate the optical performance of the pattern fork grating, we use the optical setup in Supplementary Fig. 8 to measure the diffraction patterns at different wavelengths. Light from the output fiber of a supercontinuum laser with an acoustic-optic filter is collimated to illuminate the fork grating after passing through an iris aperture, which is used to match the size of the fork grating. The diffracted light with different orders is collected by using a lens and then projected onto a CCD camera. The recorded diffraction patterns are shown in Fig. 4f of the main text. These experimental intensity profiles at the different wavelength show the expected doughnut-shape features with a dark center in the ±1 orders, which implies the creation of optical vortices and therefore confirms the validity of the fork grating patterned with our holographic lithography. In addition, the power of the zero-order diffraction changes with the illuminating wavelength, which is caused by

the wavelength-dependent phase modulation for such a binary fork grating. According to the thickness of the binary fork grating and the refractive index of the photoresist, we have doubly confirmed that both the simulated and experimental results are highly consistent, thus verifying the high quality of our pattern fork grating.